\documentclass[twocolumn,aps,pr,superscriptaddress,preprintnumbers,nofootinbib,10pt]{revtex4-2}
\usepackage{amsmath,amssymb}
\usepackage[dvipdf,dvips]{graphicx}
\usepackage{color}
\usepackage{hyperref}
\usepackage{url}
\usepackage{slashed}
\usepackage{subfigure}
\usepackage{amsmath}
\usepackage{amsfonts}
\usepackage{float} 
\usepackage{amssymb}
\usepackage{epsfig}
\usepackage{graphics}
\usepackage{euscript}
\usepackage{slashed}
\usepackage{epstopdf}
\usepackage[utf8]{inputenc}
\allowdisplaybreaks
\usepackage{pifont}
\usepackage{dsfont}
\usepackage{MnSymbol}
\usepackage{verbatim}
\usepackage{graphicx}
\usepackage{latexsym}
\usepackage{courier}

\usepackage{tikz-feynman}

\hypersetup{
colorlinks=true,
citecolor=blue,
citebordercolor=red,
linktoc=all,
linkcolor=blue,
urlcolor=blue
}

\def \and{\textmd{and}}

\newcommand{\ScalarTad}{\ensuremath{\Sigma_\phi^\textmd{tad}}{}}
\newcommand{\GluonTad}{\ensuremath{\Sigma_A^\textmd{tad}}{}}
\newcommand{\MixedSun}{\ensuremath{\Sigma_{\phi A}^\textmd{sun}}{}}
\newcommand{\ScalarTadFin}{\ensuremath{\Sigma_{\phi,\textmd{fin}}^\textmd{tad}}{}}
\newcommand{\GluonTadFin}{\ensuremath{\Sigma_{A,\textmd{fin}}^\textmd{tad}}{}}
\newcommand{\MixedSunFin}{\ensuremath{\Sigma_{\phi A,\textmd{fin}}^\textmd{sun}}{}}
\newcommand{\CounterTerm}{\ensuremath{\Gamma^{(2)}_\textmd{c.t.}}{}}

\def \be{\begin{equation}}
\def \ee{\end{equation}}

\def \bea{\begin{eqnarray}}
\def \eea{\end{eqnarray}}

\graphicspath{{./}}
\newbox{\ORCIDicon}
\sbox{\ORCIDicon}{\large\includegraphics[width=0.8em]{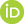}}

\begin{document}

\title{Refined Gribov-Zwanziger theory coupled to scalar fields in the Landau gauge}

\author{Gustavo P. de Brito\,\href{https://orcid.org/0000-0003-2240-528X}{\protect\usebox{\ORCIDicon}}} \email{gustavo@cp3.sdu.dk}
\affiliation{CP3-Origins, University of Southern Denmark, Campusvej 55, DK-5230 Odense M, Denmark}
\author{Philipe De Fabritiis\,\href{https://orcid.org/0000-0001-5455-6889}{\protect\usebox{\ORCIDicon}}}\email{pdf321@cbpf.br}
\affiliation{CBPF $-$ Centro Brasileiro de Pesquisas F\'isicas, Rua Dr. Xavier Sigaud 150, 22290-180, Rio de Janeiro, RJ, Brazil}
\author{Antonio D. Pereira\,\href{https://orcid.org/0000-0002-6952-2961}{\protect\usebox{\ORCIDicon}}} \email{adpjunior@id.uff.br}
\affiliation{Institute for Mathematics, Astrophysics and Particle Physics (IMAPP), Radboud University, Heyendaalseweg 135, 6525 AJ Nijmegen, The Netherlands}
\affiliation{Instituto de F\'isica, Universidade Federal Fluminense, Campus da Praia Vermelha, Av. Litor\^anea s/n, 24210-346, Niter\'oi, RJ, Brazil}

\begin{abstract}
The Refined Gribov-Zwanziger (RGZ) action in the Landau gauge accounts for the existence of infinitesimal Gribov copies as well as the dynamical formation of condensates in the infrared of Euclidean Yang-Mills theories. We couple scalar fields to the RGZ action and compute the one-loop scalar propagator in the adjoint representation of the gauge group. We compare our findings with existing lattice data.  The fate of BRST symmetry in this model is discussed, and we provide a comparison to a previous proposal for a non-minimal coupling between matter and the RGZ action. We find good agreement with the lattice data of the scalar propagator for the values of the mass parameters that fit the RGZ gluon propagator to the lattice. This suggests that the non-perturbative information carried by the gluon propagator in the RGZ framework provides a suitable mechanism to reproduce the behavior of correlation functions of colored matter fields in the infrared.
\end{abstract}

\maketitle

\section{Introduction \label{Sec:Intro}}

Understanding the mechanism that drives color confinement in Yang-Mills (YM) theories is one of the most challenging open problems in quantum field theory. As opposed to its asymptotic freedom in the ultraviolet \cite{Politzer:1973fx,Gross:1973id}, YM theories become strongly coupled in the infrared and perturbation theory breaks down. Different aspects of the confining nature of YM theories can be understood by different non-perturbative or effective descriptions and, hopefully, the synergy between such approaches will provide a complete and consistent understanding of the mechanism behind confinement. For a collection of such different perspectives, see, e.g., \cite{Greensite:2011zz,Brambilla:2014jmp}.

Treating YM theories within the framework of continuum quantum field theory typically requires the introduction of a gauge fixing condition. From a path-integral perspective, this is usually achieved by the so-called Faddeev-Popov (FP) procedure \cite{Faddeev:1967fc}. Although very successful in the perturbative regime, the usual assumptions in the FP procedure do not hold beyond that. This was identified in the Landau gauge by Gribov in \cite{Gribov:1977wm}. It was shown to exist field configurations that satisfy the Landau gauge condition and that are connected by gauge transformations. Such configurations are known as Gribov or gauge copies and their existence is what is known in the literature as the Gribov problem. In \cite{Singer:1978dk}, Singer showed that this is not a particular shortcome of the Landau gauge, but rather a generic feature of global gauge fixings in field space. The existence of Gribov copies violates one of the assumptions of the FP procedure since the gauge-fixing condition does not select a unique representative per gauge orbit. This suggests a modification of the gauge-fixing procedure in order to remove the Gribov copies. Up to date, such an improvement was successfully implemented only for infinitesimal Gribov copies, i.e., those generated by infinitesimal gauge transformations in the Landau gauge. The central idea behind the upgrade of the gauge-fixing procedure corresponds to, on top of the standard FP method, implement a restriction of the path integral to a region $\Omega$ which is free of infinitesimal Gribov copies. Such a region $\Omega$ is known as the Gribov region. It is bounded in all directions in field space, convex, and all gauge orbits cross it at least once \cite{DellAntonio:1991mms}. Such non-trivial properties ensure that the region $\Omega$ is a suitable candidate to restrict the path integral to. The boundary $\partial\Omega$ of the Gribov region is known as the Gribov horizon. The restriction of the functional measure to $\Omega$ was worked out at leading order in \cite{Gribov:1977wm} and generalized to all orders in \cite{Zwanziger:1989mf} by means of a different method. Extending the procedure of \cite{Gribov:1977wm} up to all orders in a perturbative expansion leads to the same result as in \cite{Zwanziger:1989mf} as demonstrated in \cite{Capri:2012wx}. Aftermath, as explained in \cite{Zwanziger:1989mf}, the restriction to $\Omega$ can be achieved by the introduction of an effective term into the Boltzmann weight of the partition function of the gauge-fixed Euclidean YM theories in the Landau gauge. Such a term is known as the horizon function and is non-local. Together with the horizon function, a mass paramater known as the Gribov parameter is introduced. It is not a free parameter, but fixed in a self-consistent way by a gap equation. See \cite{Vandersickel:2012tz} for a detailed discussion about the derivation of the horizon function and \cite{Sobreiro:2005ec} for a pedagogical introduction to the Gribov problem. The resulting non-local action is known as the Gribov-Zwanziger (GZ) action in the Landau gauge. Remarkably, the non-locality can be tamed by the introduction of suitable auxiliary fields, rendering an action that is local and renormalizable at all orders in perturbation theory \cite{Zwanziger:1989mf,Vandersickel:2012tz}. Either in non-local or local forms, 
we will refer to this action simply as the GZ action. Thus, the GZ action in the Landau gauge implements the restriction of the path integral to the Gribov region in a local and renormalizable way. The gluon propagator arising from the GZ action vanishes at vanishing momentum and violates reflection posivity. At tree-level, the gluon propagator features complex conjugate poles. This hampers the interpretation of the gluon as a physical excitation in the spectrum, hinting towards confinement. The ghost propagator is enhanced in the infrared, characterizing what is known as \textit{scaling} behavior. 

In \cite{Dudal:2007cw,Dudal:2008sp}, it was pointed out that the GZ action suffers from infrared instabilities leading to the formation of dimension-two condensates. In particular, the auxiliary fields introduced to localize the GZ action acquire their own dynamics and give rise to a condensate. The inclusion of the gluon and auxiliary fields condensates leads to a new action known as the Refined Gribov-Zwanziger (RGZ) action. The accompanying masses of the condensates are fixed by their own gap equations and, thus, are not free. The gluon propagator arising from the RGZ action attains a non-vanishing value at vanishing momentum and the FP ghost propagator is not enhanced in the infrared. Such a behavior is known as \textit{massive} or \textit{decoupling} solution. The tree-level propagator fits very well the lattice data in the infrared see, e.g., \cite{Cucchieri:2007md,Cucchieri:2007rg,Sternbeck:2007ug,Maas:2008ri,Bogolubsky:2009dc,Bornyakov:2009ug,Bornyakov:2013pha,Cucchieri:2011ig,Oliveira:2012eh,Maas:2015nva,Duarte:2016iko,Dudal:2018cli}. Lattice simulations display a non-vanishing value for the gluon propagator in the deep infrared. Therefore, the RGZ action provides a local and renormalizable framework which accounts for the existence of infinitesimal Gribov copies as well as the dynamical formation of condensates in the infrared. Moreover, it provides propagators for the gluon and FP ghosts that are in agreement with lattice data. It has been used for the computation of glueball masses \cite{Dudal:2009zh,Dudal:2010cd} which compare well with lattice data and provides the correct sign for the Casimir energy in the MIT bag model \cite{Canfora:2013zna}. Thermodynamic properties were investigated in, e.g., \cite{Canfora:2013kma,Canfora:2015yia}. 

A remarkable feature of the (R)GZ action in the Landau gauge proposed in \cite{Dudal:2008sp} is that it breaks BRST symmetry in an explicit but soft way. The consequences of such a breaking were deeply investigated over the past decades \cite{Maggiore:1993wq,Baulieu:2008fy,Dudal:2009xh,Sorella:2009vt,Sorella:2010it,Capri:2010hb,Dudal:2012sb,Serreau:2012cg,Serreau:2013ila,Lavrov:2011wb,Lavrov:2013boa,Moshin:2014xka,Pereira:2013aza,Pereira:2014apa,Cucchieri:2014via,Schaden:2014bea,Schaden:2015uua}. Recently, it was proposed a BRST-invariant formulation of the (R)GZ action \cite{Capri:2015ixa} thanks to the introduction of a dressed, gauge-invariant field $A^h_\mu$, see, e.g., \cite{Zwanziger:1990tn,Lavelle:1995ty}. The new BRST-symmetric formulation enabled the extension of the RGZ action to linear covariant and Curci-Ferrari gauges in harmony with gauge-parameter dependence control \cite{Capri:2015nzw,Pereira:2016fpn,Capri:2016aqq,Capri:2016gut,Capri:2017bfd,Capri:2018ijg}.

Since the Gribov problem affects directly the pure gauge sector and taking into account the existence of infinitesimal Gribov copies gives rise to a picture that seems to be compatible with the infrared behavior of at least the two-point functions of gluon and FP ghosts, a natural question is to understand how colored matter is coupled to it. An important issue to be addressed is whether the removal of Gribov copies in the gluon sector can have a dynamical  repercussion leading to a consistent picture of quark confinement. In particular, one can ask if the coupling with matter requires or not a non-minimal modification or if the standard minimal coupling between gauge fields and matter leads to a consistent picture. It is clear that if matter fields are minimally coupled to the RGZ action, then the tree-level matter propagators are just the standard ones. Hence, in this approach, any influence of the elimination of Gribov copies requires loop corrections. In this work, we give a step forward on that and investigate the one-loop propagator of scalar fields in the adjoint representation of the gauge group. Alternatively, a non-minimal coupling between matter and the RGZ action was proposed in \cite{Capri:2014bsa} and further explored in \cite{Capri:2017abz}. It provides tree-level propagators for scalars and quarks which are in qualitative agreement with lattice data. The drawback is that this requires the introduction of a new set of mass parameters by hand and a multitude of extra fields. It is not clear whether such a non-minimal coupling can be seen as an effective way of taking into account loop effects or if it is genuinely required in the formalism. This work aims at providing first directions on this question.

This paper is organized as follows: In Sect.~\ref{Sec:RGZOverview} we provide a short review of the Refined Gribov-Zwanziger framework to fix notation and point out the relevant aspects of this theory for the purposes of this work. In Sect.~\ref{Sec:CSRGZ} we discuss two different prescriptions to couple scalar fields to the Refined Gribov-Zwanziger action. The status of the BRST symmetry in the RGZ framework without and with the coupling of scalar fields is discussed in Sect.~\ref{Sec:BRSTSym} and Sect.~\ref{Sec:RGZScalarBRST}. The special role played by the Landau gauge in practical computations is outlined in Sect.~\ref{Sec:RGZScalarBRST}. The computation of the one-loop scalar-propagator in the RGZ minimally coupled to scalar fields is reported in Sect.~\ref{Sec:SFPMS} and our findings are compared with the lattice data reported in \cite{Maas:2018sqz} in Sect.~\ref{Sec:CompLattice}. Sects.~\ref{Sec:Positivity} and \ref{Sec:CompMSNPS} contain discussions on the fate of reflection positivity and the non-minimal matter coupling respectively. After that we present our conclusions in Sect. \ref{Sec:Conclusions} and collect relevant conventions in an Appendix.

\section{Brief overview on the Refined Gribov-Zwanziger action \label{Sec:RGZOverview}}

Consider YM theories in $d$ Euclidean dimensions with gauge group $SU(N)$ quantized in the Landau gauge, i.e., $\partial_\mu A^a_\mu = 0$ is the gauge-fixing condition and $a=1,\ldots ,N^2-1$. Employing the FP procedure, the gauge fixed partition function is written as
\begin{equation}
\EuScript{Z}_{\rm FP} = \int[\EuScript{D}\mu]_{\rm FP}\,{\rm e}^{-S_{\rm YM}- S_{\rm FP}}\,,
\label{Sect:Overview1}
\end{equation}
with $[\EuScript{D}\mu]_{\rm FP} = [\EuScript{D}A][\EuScript{D}b][\EuScript{D}\bar{c}][\EuScript{D}c]$ and\footnote{The short-hand notation $\int{\rm d}^dx = \int_{x^d}$ is employed. For $d=4$, we simply write $\int{\rm d}^4x = \int_{x}$.}
\begin{equation}
S_{\rm YM} = \frac{1}{4}\int_{x^d} F^{a}_{\mu\nu}F^{a}_{\mu\nu}\,,
\label{Sect:Overview2}
\end{equation}
and
\begin{equation}
S_{\rm FP} = \int_{x^d} \left(ib^a\partial_\mu A^a_\mu - \bar{c}^a\EuScript{M}^{ab}(A)c^b\right)\,,
\label{Sect:Overview3}
\end{equation}
with $F^{a}_{\mu\nu} = \partial_\mu A^a_\nu-\partial_\nu A^a_\mu + gf^{abc}A^b_\mu A^c_\nu$ being the field strength. The operator $\EuScript{M}^{ab} (A) = -\partial_\mu D^{ab}_\mu$ is the FP operator and $D^{ab}_{\mu}= \delta^{ab}\partial_\mu - gf^{abc}A^{c}_\mu$ is the covariant derivative in the adjoint representation of the gauge group.

According to Gribov (and Singer) \cite{Gribov:1977wm,Singer:1978dk}, the partition function \eqref{Sect:Overview1} still sums over spurious configurations, i.e, over field configurations that satisfy the Landau gauge condition and are connected by gauge transformations. Those configurations, the so-called Gribov(-Singer) copies, must be eliminated from \eqref{Sect:Overview1} by a suitable improvement of the FP procedure. The elimination can be achieved by means of the restriction of the path integral to a region free of copies, the so-called Fundamental Modular Region (FMR). Nevertheless, such a restriction is very difficult to be implemented and unknown to this date. A more modest attempt but already very non-trivial was proposed by Gribov in \cite{Gribov:1977wm} and improved by Zwanziger in \cite{Zwanziger:1989mf}. Essentially, it was proposed to eliminate those copies that are generated by infinitesimal gauge transformations, i.e., the infinitesimal Gribov copies. In the Landau gauge, this is achieved by defining the Gribov region $\Omega$ by
\begin{equation}
\Omega = \left\{A^a_\mu\,,\,\partial_\mu A^a_\mu = 0\, |\, \EuScript{M}^{ab}> 0 \right\}\,,
\label{Sect:Overview4}
\end{equation}
and imposing the restriction of \eqref{Sect:Overview1} to $\Omega$. The Gribov region is defined by the positivity of the FP operator which is Hermitian in the Landau gauge. Moreover, it features important geometrical features as: it is bounded in every direction; it is convex; all gauge orbits cross $\Omega$ at least once \cite{DellAntonio:1991mms}. This ensures that restricting the path integral to $\Omega$ does not leave out any physical configuration. The boundary $\partial \Omega$ of such a region is known as the Gribov horizon. Yet the Gribov region is not free of Gribov copies but just the infinitesimal ones \cite{vanBaal:1991zw}. Copies generated by finite gauge transformations are eliminated just by a further restriction to the FMR. Effectively, the path integral is restricted to the Gribov region by a modification of the Boltzmann weight as
\begin{eqnarray}
\EuScript{Z}_{\rm GZ} &=& \int_{\Omega} [\EuScript{D}\mu]_{\rm FP}\,{\rm e}^{-S_{\rm YM}- S_{\rm FP}}\nonumber\\
&=&\int[\EuScript{D}\mu]_{\rm FP}\,{\rm e}^{-S_{\rm YM}- S_{\rm FP}-\gamma^4H (A)+d\gamma^4(N^2-1)V}\,.
\label{Sect:Overview5}
\end{eqnarray}
In \eqref{Sect:Overview5}, $V$ represents the spacetime volume and $d$ its dimensionality. The function $H(A)$ is the so-called Horizon function and it is expressed as
\begin{equation}
H(A) = g^2 \int_{x^d,y^d} f^{abc}A^{b}_\mu (x) \Big[\EuScript{M}^{-1}\Big]^{ad}(x,y)f^{dec}A^e_\mu (y)\,.
\label{Sect:Overview6}
\end{equation}
The parameter $\gamma$ is a massive parameter known as the Gribov parameter. It is not free but fixed by a gap equation given by
\begin{equation}
\langle H(A) \rangle = d V(N^2-1)\,,
\label{Sect:Overview7}
\end{equation}
with $\langle \ldots \rangle$ being computed with the measure defined by \eqref{Sect:Overview5}. 

The effective restriction to the Gribov region $\Omega$ amounts to introduce the Horizon function which is non-local due to the presence of the inverse of the FP operator $\EuScript{M}$. Remarkably, it can be cast in local form by the introduction of suitable auxiliary fields, namely, a pair of commuting fields $(\bar{\varphi},\varphi)^{ab}_{\mu}$ and a pair of Grassmannian ones $(\bar{\omega},\omega)^{ab}_\mu$. In terms of the localizing fields, the action which effectively implements the restriction of the path integral to $\Omega$ is
\begin{equation}
S_{\rm GZ} = S_{\rm YM} + S_{\rm FP} + S_H\,,
\label{Sect:Overview8}
\end{equation}
with
\begin{eqnarray}
S_H &=& \int_{x^d}\Big(\bar{\varphi}^{ab}_\mu\,\EuScript{M}^{ac}(A)\,\varphi^{cb}_\mu-\bar{\omega}^{ab}_\mu\,\EuScript{M}^{ac}(A)\,\omega^{cb}_\mu\Big)\nonumber\\
&+&ig\gamma^2 \int_{x^d} f^{abc}A^a_\mu (\varphi^{bc}_\mu + \bar{\varphi}^{bc}_\mu)\,.
\label{Sect:Overview9}
\end{eqnarray}
The action \eqref{Sect:Overview8} is known as the Gribov-Zwanziger (GZ) action. It is local, renormalizable at all orders in perturbation theory, and effectively implements the restriction of the path integral to the Gribov region \cite{Zwanziger:1989mf}. In the presence of the new auxiliary localizing fields, the partition function $\EuScript{Z}_{\rm GZ}$ is expressed as
\begin{equation}
\EuScript{Z}_{\rm GZ} = \int [\EuScript{D}\mu]_{\rm GZ}\,{\rm e}^{-S_{\rm GZ}+d\gamma^4(N^2-1)V}\,,
\label{Sect:Overview10}
\end{equation}
with $[\EuScript{D}\mu]_{\rm GZ} = [\EuScript{D}A][\EuScript{D}b][\EuScript{D}\bar{c}][\EuScript{D}c][\EuScript{D}{\bar \varphi}][\EuScript{D}\varphi][\EuScript{D}\bar{\omega}][\EuScript{D}\omega]$. In this local formulation, the gap equation that fixes the Gribov parameter is expressed as
\begin{equation}
\frac{\partial \EuScript{E}_v}{\partial \gamma^2}\Bigg|_{\gamma^2 \neq 0} = 0\,,
\label{Sect:Overview11}
\end{equation}
where $\EuScript{E}_v$ is the vacuum energy defined by ${\rm e}^{-V\EuScript{E_v}} = \EuScript{Z}_{\rm GZ}$.

The GZ action leads to two striking features: the gluon propagator vanishes exactly at vanishing momentum while the FP ghost propagator is enhanced in the deep infrared, i.e., it behaves as $\sim 1/p^4$, with $p$ denoting the Euclidean momentum. Such properties are in qualitative agreement with the so-called scaling solutions for the non-perturbative propagators of pure YM theories see, e.g., \cite{vonSmekal:1997ohs,Fischer:2008uz}. Lattice simulations performed with bigger lattices revealed two distinct features: the gluon propagator did not attain vanishing value at vanishing momentum and the ghost propagator was not enhanced in the deep infrared in $d=3,4$ while the scaling-type solution persisted in $d=2$. Such a new behavior was reported in, e.g., \cite{Cucchieri:2007rg} and is dubbed as massive or decoupling solution. 

The coexistence of the GZ scenario with the massive/decoupling solution came in \cite{Dudal:2008sp} where it was identified that suitable infrared instabilities must be taken into account in the theory. In particular, the auxiliary fields introduced to localize the Horizon function develop their own dynamics and give rise to condensates in $d>2$, see \cite{Dudal:2008xd,Dudal:2008rm,Dudal:2011gd}. The inclusion of such lower-dimensional operators (both of auxiliary fields and gluons) in the GZ action gave birth to the so-called Refined Gribov-Zwanziger (RGZ) scenario \cite{Dudal:2008sp}. Hence, the action is written as
\begin{equation}
S_{\rm RGZ} = S_{\rm GZ} + S_{\rm cond}
\label{Sect:Overview12}
\end{equation} 
with
\begin{equation}
S_{\rm cond} = \int_{x^d}\left[\frac{m^2}{2}A^a_\mu A^a_\mu+M^2(\bar{\varphi}^{ab}_\mu \varphi^{ab}_\mu - \bar{\omega}^{ab}_\mu \omega^{ab}_\mu)\right]\,.
\label{Sect:Overview13}
\end{equation}
The RGZ action leads to a tree-level gluon propagator that attains a non-vanishing value at zero momentum and a ghost propagator that is not enhanced in the deep infrared, i.e., it goes as $\sim 1/p^2$. In the next Section, we will discuss two different ways of coupling scalar fields to this theory.

\section{Coupling Scalar Fields to the RGZ action \label{Sec:CSRGZ}}

The RGZ action is a promising candidate to describe the infrared dynamics of pure YM theories. Yet it remains to be understood how matter should be coupled to this theory. The introduction of matter fields in the standard way, i.e., as one would introduce to the standard YM action (we shall refer to this as minimal scheme) immediately leads to the fact that the tree-level propagator of matter fields will be completely blind to the non-perturbative effects introduced by the Gribov horizon and the condensates. This picture is consistent with the fact that the Gribov copies will engender a modification in the pure gauge sector and the effects on the matter dynamics will come in by taking into account quantum fluctuations. It is logical to expect that the non-perturbative behavior of matter fields will then require the application of non-perturbative techniques such as the Functional Renormalization Group or Dyson-Schwinger equations to the RGZ action coupled do matter. Although this is a completely legitimate path, it is technically challenging due to the complicated structure of the RGZ action. Instead, another possibility to be explored is to compute quantum corrections to matter-fields propagators within perturbation theory on top of the RGZ framework. Since the gluon propagator carries non-perturbative information already at tree-level, this is an attempt worth exploring. As a further motivation to this, this type of strategy was employed in effective models such as the massive (Curci-Ferrari) model for the quark sector \cite{Pelaez:2014mxa,Pelaez:2017bhh} leading to promising results.

Alternatively, one can propose a non-minimal coupling between matter and gauge fields in the RGZ scenario as done in \cite{Capri:2014bsa,Capri:2017abz}. The central idea is that, within the Gribov horizon, the quantity $\EuScript{M}^{-1}$ is well-defined and it would naturally couple to colored fields (we will refer to this as non-minimal scheme). In this case, as we shall review below, the tree-level propagator of the matter field displays non-perturbative properties which fit well lattice data, see \cite{Capri:2014bsa}. The drawback of this approach, however, is that it introduces new massive parameters that do not have a clear geometrical picture as in the pure-gauge sector. The line of research set out in this paper aims at providing a clearer direction to which scheme of matter coupling to the RGZ action must be employed. In the following subsections, we make a brief technical overview of both schemes using scalar fields in the adjoint representation as our prototype.  

\subsection{Minimal Scheme \label{SubSect:MS}}
The RGZ-scalar system in the minimal scheme is defined by the action
\begin{equation}
S^{\rm ms}_{\phi \rm RGZ} = S_{\rm RGZ} + S_{\phi}\,,
\label{SubSect:MS.1}
\end{equation}
with
\begin{equation}
S_\phi = \int_{x^d}\left[\frac{1}{2}(D^{ab}_\mu \phi^b)(D^{ac}_\mu \phi^c)+\frac{m^2_\phi}{2}\phi^a \phi^a + \frac{\lambda}{4!}(\phi^a \phi^a )^2\right]\,.
\label{SubSect:MS.2}
\end{equation}
In this case, the tree-level scalar-field propagator is just the standard one, i.e., 
\begin{equation}
\langle \phi^a (p) \phi^b (-p)\rangle_0 = \frac{\delta^{ab}}{p^2+m^2_\phi}\,.
\label{SubSect:MS.3}
\end{equation}
Clearly, the tree-level propagator does not feel the presence of the Gribov horizon, which would only enter in the loop contributions. Before discussing more about that, we introduce the non-minimal scheme in the following Subsection.

\subsection{Non-minimal Scheme \label{SubSect:NPS}}

In this scheme, a Horizon-like function is introduced for the scalar fields, i.e., one introduces the function $\mathcal{H}(\phi)$ defined by
\begin{equation}
\mathcal{H}(\phi) = g^2 \int_{x^d,y^d} f^{abc}\phi^{b} (x) \,\Big[\EuScript{M}^{-1}\Big]^{ad}(x,y)f^{dec}\phi^e (y)\,.
\label{SubSect:NPS.1}
\end{equation}
Hence, the modified scalar action accounting for the non-minimal coupling is replaced by 
\begin{equation}
S_{\phi} \to S_\phi + \sigma^4\mathcal{H}(\phi)\,,
\label{SubSect:NPS.2}
\end{equation}
with $\sigma^4$ being a mass-like parameter playing the analogue role of the Gribov parameter $\gamma$. One of the drawbacks of the non-minimal scheme is that, unlike the Gribov parameter, the parameter $\sigma$ does not have a geometrical interpretation unless the $d$-dimensional theory arises from a dynamical reduction from the pure RGZ action in $d+1$ dimensions, see \cite{Guimaraes:2016okb}. The Horizon-like function $\mathcal{H}(\phi)$ is non-local and can be cast in local form in analogy to the localization procedure in the GZ action. Thus, 
\begin{eqnarray}
\sigma^4 \mathcal{H}(\phi)&\to& \int_{x^d}\Big(\bar{\zeta}^{ab} \EuScript{M}^{ac}(A)\zeta^{cb}-\bar{\theta}^{ab}\EuScript{M}^{ac}(A)\theta^{cb}\Big)\nonumber\\
&+&ig\sigma^2\int_{x^d}f^{abc}\phi^{a}(\bar{\zeta}+\zeta)^{bc}\,,
\label{SubSect:NPS.3}
\end{eqnarray}
where $(\bar{\zeta},\zeta)^{ab}$ are commuting fields and $(\bar{\theta},\theta)^{ab}$ anti-commuting ones. Similarly to the refinement of the GZ action, the auxiliary fields just introduced acquire their own dynamics and generate condensates, at least in $d>2$, see \cite{Capri:2017abz}. We introduce to \eqref{SubSect:NPS.3}, the term
\begin{equation}
S^{\rm cond}_{\phi} = M^2_\phi \int_{x^d} (\bar{\zeta}^{ab}{\zeta}^{ab}-\bar{\theta}^{ab}{\theta}^{ab})\,.
\label{SubSect:NPS.4}
\end{equation}
Finally, the scalar-field action in the non-minimal scheme is written as
\begin{eqnarray}
S^{\rm nm}_{\phi} &=& S_{\phi} + \int_{x^d}\Big(\bar{\zeta}^{ab} \EuScript{M}^{ac}(A)\zeta^{cb}-\bar{\theta}^{ab}\EuScript{M}^{ac}(A)\theta^{cb}\Big)\nonumber\\
&+&ig\sigma^2\int_{x^d}f^{abc}\phi^{a}(\bar{\zeta}+\zeta)^{bc}+S^{\rm cond}_{\phi}\,.
\label{SubSect:NPS.5}
\end{eqnarray}
Therefore, the RGZ-scalar system in the non-minimal scheme is defined by
\begin{equation}
S^{\rm nm}_{\phi {\rm RGZ}} = S_{\rm RGZ}+ S^{\rm nm}_{\phi}\,.
\label{SubSect:NPS.6}
\end{equation}
The tree-level propagator of the scalar field in the non-minimal scheme is given by
\begin{equation}
\langle \phi^a (p) \phi^b (-p)\rangle_0 = \delta^{ab}\frac{p^2+M^2_\phi}{(p^2+m^2_\phi)(p^2+M^2_\phi)+2Ng^2\sigma^4}\,,
\label{SubSect:NPS.7}
\end{equation}
which fits well the lattice data, as discussed in \cite{Capri:2014bsa}.

In the following, we will discuss formal aspects of the RGZ-scalar system with a focus towards the minimal scheme. In particular, we will compute the one-loop correction to the two-point function of scalar fields in the minimal scheme. When possible, we provide a comparison between both frameworks. An important issue to be addressed is the fate of the BRST symmetry in the RGZ-scalar system. This is the topic of the next Section.

\section{BRST symmetry in the RGZ Framework\label{Sec:BRSTSym}}

\subsection{Soft breaking of BRST symmetry\label{SubSec:BRSTBreakSym}}

An important outcome of the FP procedure is the so-called BRST symmetry \cite{Becchi:1975nq,Tyutin:1975qk,Baulieu:1981sb}. In particular, the gauge-fixed action given by \eqref{Sect:Overview2} and \eqref{Sect:Overview3} is invariant under the transformations,
\begin{eqnarray}
\mathsf{s}A^a_\mu &=& - D^{ab}_\mu c^b\,,\nonumber\\
\mathsf{s}c^a &=& \frac{g}{2}f^{abc}c^b c^c\,,\nonumber\\
\mathsf{s}\bar{c}^a &=& ib^a\,,\nonumber\\
\mathsf{s}b^a &=& 0\,.
\label{Sect:BRSTSym.1}
\end{eqnarray}
The BRST operator $\mathsf{s}$ has ghost number one and is nilpotent, i.e., $\mathsf{s}^2=0$. The RGZ action can be expressed as,
\begin{eqnarray}
S_{\rm RGZ} &=& S_{\rm YM} + S_{\rm FP} + \mathsf{s}\int_{x^d}\bar{\omega}^{ab}_\mu\EuScript{M}^{ac}(A)\varphi^{cb}_\mu \nonumber\\
&+&ig\gamma^2\int_{x^d}f^{abc}A^a_\mu (\varphi^{bc}_{\mu}+\bar{\varphi}^{bc}_{\mu})+S_{\rm cond}\,,
\label{Sect:BRSTSym.2}
\end{eqnarray}
with 
\begin{eqnarray}
\mathsf{s}\bar{\omega}^{ab}_\mu = \varphi^{ab}_\mu\,,\qquad \mathsf{s}\varphi^{ab}_\mu = 0\,,\nonumber\\
\mathsf{s}\bar{\varphi}^{ab}_\mu = \omega^{ab}_\mu\,,\qquad \mathsf{s}\omega^{ab}_\mu = 0\,.
\label{Sec:BRSTSym.3}
\end{eqnarray}
Two comments are in order: The auxiliary fields are introduced as BRST-doublets and therefore do not affect the non-trivial part of the cohomology of the BRST operator $\mathsf{s}$. Part of the localized Horizon function can be written as a BRST-exact term and by expanding such a term, one gets an extra term with respect to \eqref{Sect:Overview9}. However, such an extra term can be eliminated by a harmless field redefinition on the $(\bar{\omega},\omega)$ sector. From \eqref{Sect:BRSTSym.2}, it is easy to check that $\mathsf{s}S_{\rm RGZ}\neq 0$, i.e., the RGZ action breaks BRST invariance. There are two sorts of breaking: one coming from the gluon condensate and the other coming from the $\gamma$-dependent contribution. The former is less dangerous because it is BRST-invariant on-shell in the Landau gauge while the latter consists to a genuine breaking. Besides being an explicit breaking of the BRST symmetry, it is soft, i.e., it is proportional to the mass parameter $\gamma^2$. In the deep ultraviolet, $\gamma^2\to 0$ and BRST-invariance is recovered as it should. Such a breaking was explored in great detail over the last years, see, e.g., \cite{Maggiore:1993wq,Baulieu:2008fy,Dudal:2009xh,Sorella:2009vt,Sorella:2010it,Capri:2010hb,Dudal:2012sb,Serreau:2012cg,Serreau:2013ila,Lavrov:2011wb,Lavrov:2013boa,Moshin:2014xka,Pereira:2013aza,Pereira:2014apa,Cucchieri:2014via,Schaden:2014bea,Schaden:2015uua}. In particular,
\begin{equation}
\frac{\partial S_{\rm RGZ}}{\partial \gamma^2} \neq \mathsf{s}\Delta\,,
\label{Sec:BRSTSym.4}
\end{equation}
for an insertion $\Delta$ with ghost-number $-1$. Eq.\eqref{Sec:BRSTSym.4} shows that the Gribov parameter is not akin to a gauge parameter and can enter gauge-invariant correlation functions. 

In the minimal scheme of the matter coupling, the situation is the same as the one just described, i.e., the RGZ action displays a soft breaking while the matter action is invariant under BRST symmetry, with
\begin{equation}
\mathsf{s}\phi^a = -gf^{abc}\phi^b c^c\,.
\label{Sec:BRSTSym.5}
\end{equation}
As for the non-minimal scheme, due to the introduction of the Horizon-like function, a new source of BRST breaking is generated, i.e., 
\begin{equation}
\mathsf{s} S^{\rm np}_\phi = ig\sigma^2\int_{x^d}f^{abc}\Big(-gf^{ade}\phi^d c^e (\bar{\zeta}+\zeta)^{bc} + \phi^a\theta^{bc}\Big)\,,
\label{Sec:BRSTSym.6}
\end{equation}
with
\begin{eqnarray}
\mathsf{s}\bar{\theta}^{ab} = \zeta^{ab}\,,\qquad \mathsf{s}\zeta^{ab} = 0\,,\nonumber\\
\mathsf{s}\bar{\zeta}^{ab} = \theta^{ab}\,,\qquad \mathsf{s}\theta^{ab} = 0\,.
\label{Sec:BRSTSym.7}
\end{eqnarray}
Moreover,
\begin{equation}
\frac{\partial S^{\rm np}_{\phi {\rm RGZ}}}{\partial \sigma^2} \neq \mathsf{s}\tilde{\Delta}\,,
\label{Sec:BRSTSym.8}
\end{equation}
for an insertion $\tilde{\Delta}$ with ghost number -1. Therefore, the mass parameter $\sigma^2$ is not like a gauge parameter and can enter correlation functions of gauge-invariant operators as well. 

As long as we restrict ourselves to the Landau gauge, the BRST soft breaking does not preclude a consistent treatment of the RGZ-scalar action. Yet the very fact that one would like to be able to move to different gauges lead to a potential problem when BRST invariance is broken. It is precisely the BRST symmetry that controls gauge-parameter dependence of gauge-invariant correlators. Over the last years, a proposal to restore BRST invariance in the RGZ action was made. In the next Subsection we provide a short discussion on that and establish how correlators in the RGZ-scalar system in the BRST-broken formulation are related to those in the BRST-invariant formalism.

\subsection{Restoring the BRST symmetry\label{SubSec:BRSTRestSym}}

The key ingredient for the construction of the BRST-invariant formulation of the RGZ action is the gauge-invariant dressed gauge field $A^{h,a}_\mu$, see, e.g., \cite{Zwanziger:1990tn,Lavelle:1995ty,Capri:2015ixa}. It is constructed by the minimization of the functional $\mathcal{A}^2_{\rm min}$,
\begin{equation}
\mathcal{A}^2_{\rm min} = \underset{U}{\rm min}\,{\rm Tr}\int_{x^d}A^U_\mu A^U_\mu \, ,
\label{Sec:BRSTSym.9}
\end{equation}
along the gauge orbit. By definition,
\begin{equation}
A^U_\mu = U^\dagger A_\mu U +\frac{i}{g}U^\dagger\partial_\mu U \, ,
\label{Sec:BRSTSym.10}
\end{equation}
where $U = {\rm exp}(ig\omega^a T^a)$ and $A_\mu = A^a_\mu T^a$. The parameters $\omega^a$ are the parameters of the gauge transformation and $\left\{T^a\right\}$ are the generators of the $SU(N)$ gauge group. As discussed in \cite{Capri:2015ixa}, for a given gauge orbit characterized by a field configuration $A^a_\mu$, a local minimum $A^{h}_\mu = A^{h,a}_\mu T^a$ of \eqref{Sec:BRSTSym.9} is given by
\begin{equation}
A^{h}_{\mu} = \left(\delta_{\mu\nu}-\frac{\partial_\mu \partial_\nu}{\partial^2}\right)\chi_\nu\,,
\label{Sec:BRSTSym.11}
\end{equation}
where
\begin{eqnarray}
\chi_\nu &=& A_\nu - ig\left[\frac{1}{\partial^2}\partial A, A_\nu\right]+\frac{ig}{2}\left[\frac{1}{\partial^2}\partial A,\partial_\nu \frac{1}{\partial^2}\partial A\right]\nonumber\\
&+& \mathcal{O} (A^3)\,.
\label{Sec:BRSTSym.12}
\end{eqnarray}
It is clear that the field $A^h_\mu$ is transverse, i.e., $\partial_\mu A^h_\mu = 0$. Moreover, it is gauge-invariant order by order in the coupling $g$. Finally, Eqs.\eqref{Sec:BRSTSym.11} and \eqref{Sec:BRSTSym.12} show that it is possible to write the field $A^h_\mu$ as the gauge field $A_\mu$ plus terms that always have, at least, a divergence of the gauge field, namely, $\partial_\alpha A_\alpha$. By construction, the dressed gauge-invariant field $A^h_\mu$ is BRST invariant, i.e.,
\begin{equation}
\mathsf{s}A^{h}_{\mu}=0\,.
\label{Sec:BRSTSym.13}
\end{equation}
As discussed in \cite{Capri:2015ixa}, the Horizon function $H(A)$ can be re-expressed in terms of the gauge-invariant field $A^h_\mu$, i.e., 
\begin{equation}
H(A^h) = g^2 \int_{x^d,y^d} f^{abc}A^{h,b}_\mu \Big[\EuScript{M}^{-1}(A^h)\Big]^{ad}f^{dec}A^{h,e}_\mu\,,
\label{Sec:BRSTSym.14}
\end{equation}
where the spacetime dependence of fields and the dressed-FP operator is suppressed. The dressed Horizon function \eqref{Sec:BRSTSym.14} effectively implements the restriction of the path integral to $\Omega^h$, which is defined as
\begin{equation}
\Omega^h = \left\{A^a_\mu\,,\, \partial_\mu A^{h}_\mu = 0\,\,|\,\,\EuScript{M}(A^h)>0\right\}\,,
\label{Sec:BRSTSym.15}
\end{equation}
which, in the Landau gauge, is equivalent to $\Omega$. See \cite{Capri:2015ixa}. Therefore, the Gribov-Zwanziger action in the Landau gauge can be rewritten in terms of the dressed Horizon function $H(A^h)$ leading to a BRST-invariant action. Yet the Horizon function $H(A^h)$ has two sources of non-localities: One stems from the inverse of the operator $\EuScript{M} (\circ)$ and the other one arising from $A^h_\mu$ itself. In \cite{Capri:2016aqq,Capri:2017bfd} it was worked out the localization of such an action leading to a local and renormalizable framework that effectively implements the restriction of the path integral to $\Omega$ that is compatible with BRST symmetry. The localization procedure requires the introduction of a Stueckelberg-like field $\xi^a$ which renders a non-polynomial action. Remarkably this field completely decouples in the Landau gauge, a property that gives a special technical advantage to this gauge choice, see, e.g., the discussion raised in \cite{Capri:2018ijg}. The local and BRST-invariant GZ action is written as
\begin{equation}
S^h_{\rm GZ} = S_{\rm YM} + S_{\rm FP} + S^h_H + S_{\rm aux}\,,
\label{Sec:BRSTSym.16}
\end{equation}
with
\begin{eqnarray}
S^h_H &=& \int_{x^d}\Big(\bar{\varphi}^{ab}_\mu\,\EuScript{M}^{ac}(A^h)\,\varphi^{cb}_\mu-\bar{\omega}^{ab}_\mu\,\EuScript{M}^{ac}(A^h)\,\omega^{cb}_\mu\Big)\nonumber\\
&+&ig\gamma^2 \int_{x^d} f^{abc}(A^{h})^a_\mu (\varphi^{bc}_\mu + \bar{\varphi}^{bc}_\mu)\,,
\label{Sec:BRSTSym.17}
\end{eqnarray}
and
\begin{equation}
S_{\rm aux} = \int_{x^d}\Big(\tau^a \partial_\mu (A^{h})^a_\mu-\bar{\eta}^a\EuScript{M}^{ab}(A^h)\eta^b\Big)\,.
\label{Sec:BRSTSym.18}
\end{equation}
The composite operator $(A^h)_\mu = (A^h)^a_\mu T^a$ is defined as
\begin{equation}
(A^h)_\mu = h^\dagger A_\mu h +\frac{i}{g}h^\dagger\partial_\mu h\,,
\label{Sec:BRSTSym.19}
\end{equation}
where
\begin{equation}
h = {\rm e}^{ig \xi^a T^a}\,.
\label{Sec:BRSTSym.20}
\end{equation}
The field $\xi^a$ is a Stueckelberg-like field. The field $\tau^a$ plays the role of a Lagrange multiplier introduced to impose the transversality of the field $(A^h)_\mu$. The fields $(\bar{\eta},\eta)^a$ are ghosts introduced very much like the FP ghosts to compensate the transversality condition on $(A^h)_\mu$. The auxiliary fields $(\bar{\varphi},\varphi)^{ab}_\mu$ and $(\bar{\omega},\omega)^{ab}_\mu$ are completely analogous to those introduced in the standard GZ framework. Yet an important feature stands out in Eqs.\eqref{Sec:BRSTSym.19} and \eqref{Sec:BRSTSym.20}: all the auxiliary fields but $\xi^a$ are BRST singlets, i.e.,
\begin{eqnarray}
\mathsf{s}\Phi^I = 0\,,
\label{Sec:BRSTSym.21}
\end{eqnarray}
with $\Phi^I = \left\{\bar{\varphi},\varphi,\bar{\omega},\omega,\tau,\bar{\eta},\eta\right\}$. As for the field $\xi^a$, one has
\begin{equation}
\mathsf{s}\xi^a = g^{ab}(\xi)\xi^b\,,
\label{Sec:BRSTSym.22}
\end{equation}
with
\begin{equation}
g^{ab}(\xi) = -\delta^{ab} + \frac{g}{2}f^{abc}\xi^c - \frac{g^2}{12}f^{amr}f^{mbq}\xi^q \xi^r + \EuScript{O}(g^3)\,.
\label{Sec:BRSTSym.23}
\end{equation}
This follows from the BRST-invariance of the dressed field $A^h_\mu$ which, in turn, entail the following transformations for $h$ and $h^\dagger$,
\begin{equation}
\mathsf{s}h = -igc\,h\,,\quad {\rm and} \quad \mathsf{s}h^\dagger = igh^\dagger c\,,
\label{Sec:BRSTSym.23.1}
\end{equation}
where the matrix-notation was employed.

As for the standard GZ action, lower-dimensional condensates are formed in the BRST-invariant setting leading to a BRST-invariant refined GZ action. It reads
\begin{equation}
S^h_{\rm RGZ} = S^h_{\rm GZ} + S^h_{\rm cond}\,,
\label{Sec:BRSTSym.24}
\end{equation}
where
\begin{equation}
S^h_{\rm cond} = \int_{x^d}\left[\frac{m^2}{2}(A^h)^a_\mu (A^h)^a_\mu+M^2(\bar{\varphi}^{ab}_\mu \varphi^{ab}_\mu - \bar{\omega}^{ab}_\mu \omega^{ab}_\mu)\right]\,.
\label{Sec:BRSTSym.25}
\end{equation}
The action in \eqref{Sec:BRSTSym.25} is invariant under \eqref{Sect:BRSTSym.1}, \eqref{Sec:BRSTSym.13}, \eqref{Sec:BRSTSym.21}, and \eqref{Sec:BRSTSym.22}. The path integral associated with \eqref{Sec:BRSTSym.24} is defined by
\begin{equation}
\EuScript{Z}^h_{\rm GZ} = \int [\EuScript{D}\mu]^h_{\rm GZ}\,{\rm e}^{-S^h_{\rm RGZ}+d\gamma^4(N^2-1)V}\,,
\label{Sec:BRSTSym.26}
\end{equation}
with
\begin{equation}
[\EuScript{D}\mu]^h_{\rm GZ} = [\EuScript{D}\mu]_{\rm GZ}[\EuScript{D}\xi][\EuScript{D}\tau][\EuScript{D}\bar{\eta}][\EuScript{D}\eta]\,.
\label{Sec:BRSTSym.27}
\end{equation}
As discussed in \cite{Capri:2018ijg}, the partition function \eqref{Sec:BRSTSym.26} is equivalent to $\EuScript{Z}_{\rm GZ}$ defined with the soft-BRST broken action \eqref{Sect:Overview12}. This is a consequence of the special nature of the Landau gauge and the transversality of the dressed field $A^h_\mu$. Consequently, correlation functions of gauge-invariant operators $\EuScript{O}_n (x)$ are equivalently calculable in both settings, i.e.,
\begin{equation}
\langle \EuScript{O}_1 (x_1)\ldots \EuScript{O}_n (x_n)\rangle^h_{\rm RGZ}=\langle \EuScript{O}_1 (x_1)\ldots \EuScript{O}_n (x_n)\rangle_{\rm RGZ\,.}
\label{Sec:BRSTSym.28}
\end{equation}
The expectation value $\langle \ldots \rangle^h_{\rm RGZ}$ is taken with the measure defined in \eqref{Sec:BRSTSym.26} and with the action \eqref{Sec:BRSTSym.24} as the weight. As for $\langle \ldots \rangle_{\rm RGZ}$, the measure is taken as in \eqref{Sect:Overview10} with weight \eqref{Sect:Overview12}. In practice, it is way more economical to employ the standard RGZ action (or BRST-softly broken) due to its polynomial nature as well as your reduced field content with respect to the BRST-invariant one. Moreover, the mass parameters $\rho_i = \left\{\gamma^2,m^2,M^2\right\}$ are not like gauge parameters, i.e.,
\begin{equation}
\frac{\partial S^h_{\rm RGZ}}{\partial\rho_i}\neq \mathsf{s}\Delta_i\,,
\label{Sec:BRSTSym.29}
\end{equation}
for any insertions $\Delta_i$ with ghost number $-1$ but now with the BRST transformations defining a symmetry of the theory.

Having established a BRST-invariant reformulation of the RGZ framework and the relation with the standard setup, we tackle the issue of introducing scalar matter in harmony with BRST symmetry in the next section. In particular, we provide a gauge-invariant meaning to the scalars two-point function in the Landau gauge which will turn out to be the main object of interest in this work.

\section{RGZ-Scalar System: BRST Symmetry and the Special Role of the Landau gauge\label{Sec:RGZScalarBRST}}

\subsection{Minimal Scheme \label{SubSect:MSBRST}}

As discussed in SubSect.~\ref{SubSect:MS}, the coupling of scalar fields to the RGZ framework in the minimal scheme does not add any new source of BRST breaking. Then the action 
\begin{equation}
S^{h ,{\rm ms}}_{\phi {\rm RGZ}} = S^{h}_{\phi {\rm RGZ}} + S_\phi\,,
\label{SubSect:MSBRST.1}
\end{equation}
is BRST invariant. An important issue to be addressed is if correlation functions computed in such a framework would differ from those computed in the standard setting, i.e., defined by the action \eqref{SubSect:MS.1}. We will address such a question as a particular case of the BRST-invariant non-minimal coupling between scalars and the RGZ action to be presented in the next subsection. It turns out that the correlation functions of gauge-invariant operators computed in the minimal scheme are the same in the BRST-broken and BRST-invariant scenarios. This could be guessed by the fact that in the minimal scheme, the scalar fields do not couple directly to the localizing Stueckelberg-field and therefore should not hamper its decoupling in the Landau gauge.

\subsection{Non-minimal Scheme \label{SubSect:NPSBRST}}

In such a scheme, a Horizon-like function \eqref{SubSect:NPS.1} is introduced for the matter field $\phi^a$. This is a new source of BRST-breaking terms and since the RGZ action can be cast as BRST-invariant on its own, such a new term must be dressed in a BRST-symmetric fashion. This construction was introduced in, e.g., \cite{Capri:2017abz}. It starts by the introduction of the gauge-invariant dressed scalar field $\phi^h = \phi^{h,a}T^a$ defined by
\begin{equation}
\phi^h = h^\dagger \phi\, h \,,
\label{SubSect:NPSBRST.1}
\end{equation}
leading to
\begin{equation}
\phi^{h,a} = \phi^a + gf^{abc}\xi^b\phi^c + \EuScript{O}(\xi^2)\,.
\label{SubSect:NPSBRST.2}
\end{equation}
The combination of \eqref{Sec:BRSTSym.5} with \eqref{Sec:BRSTSym.23.1} yields
\begin{equation}
\mathsf{s}\phi^h = 0\,.
\label{SubSect:NPSBRST.3}
\end{equation}
Thus the dressed Horizon-like function for the scalar field becomes
\begin{equation}
\mathcal{H}(\phi^h) = g^2 \int_{x^d,y^d} f^{abc}\phi^{h,b} \,\Big[\EuScript{M}^{-1}(A^h)\Big]^{ad}f^{dec}\phi^{h,e} \,.
\label{SubSect:NPSBRST.4}
\end{equation}
The dressed field $\phi^h$ is local albeit non-polynomial and, by construction, the dressed Horizon-like function \eqref{SubSect:NPSBRST.4} is BRST invariant. Yet the expression \eqref{SubSect:NPSBRST.4} is non-local due to the presence of the inverse of $\EuScript{M}(A^h)$. Such a non-locality can be easily dealt with by the introduction of auxiliary local fields $(\bar{\zeta},\zeta,\bar{\theta},\theta)^{ab}$ as pointed out in SubSect.~\ref{SubSect:NPS}. Therefore, in local form, the scalar action to be added to the RGZ one in the non-minimal scheme is given by
\begin{eqnarray}
S^{h,\rm nm}_{\phi} &=& S_{\phi} + \int_{x^d}\Big(\bar{\zeta}^{ab} \EuScript{M}^{ac}(A^h)\zeta^{cb}-\bar{\theta}^{ab}\EuScript{M}^{ac}(A^h)\theta^{cb}\Big)\nonumber\\
&+&ig\sigma^2\int_{x^d}f^{abc}\phi^{h,a}(\bar{\zeta}+\zeta)^{bc}+S^{h,\rm cond}_{\phi}\,,
\label{SubSect:NPSBRST.5}
\end{eqnarray}
with
\begin{equation}
S^{h,\rm cond}_{\phi} = M^2_\phi \int_{x^d} (\bar{\zeta}^{ab}{\zeta}^{ab}-\bar{\theta}^{ab}{\theta}^{ab})\,,
\label{SubSect:NPSBRST.6}
\end{equation}
being the action introduced to account for the condensation of the auxiliary localizing fields.
Unlike the undressed formulation, the auxiliary fields introduced to localize the dressed Horizon-like function are BRST singlets, i.e.,
\begin{equation}
s\tilde{\Phi}^I = 0\,,
\label{SubSect:NPSBRST.7}
\end{equation}
with $\tilde{\Phi}^I = (\bar{\zeta},\zeta,\bar{\theta},\theta)^{ab}$. Finally, the local and BRST-invariant RGZ-Scalar action within the non-minimal scheme is defined by
\begin{equation}
S^{h,\rm nm}_{\phi {\rm RGZ}} = S^h_{\rm RGZ}+ S^{h,\rm nm}_{\phi}\,.
\label{SubSect:NPSBRST.8}
\end{equation}
Two comments are in order. Firstly, by setting the mass parameter $\sigma^2$ to zero, it is straightforward to check that upon integration of the auxiliary fields $\tilde{\Phi}^I$, one recovers the RGZ-scalar system in the minimal scheme. This can be formally translated into
\begin{equation}
S^{h ,{\rm ms}}_{\phi {\rm RGZ}} = S^{h ,{\rm nm}}_{\phi {\rm RGZ}}\Big|_{\sigma^2 = 0}\,.
\label{SubSect:NPBRST.9}
\end{equation}
Secondly, the mass parameters $\sigma^2$ and $M^2_\phi$ are not like gauge parameters since they are coupled to BRST-closed terms. Hence, they can be present in correlation functions of gauge-invariant operators. Finally, the path integral of the RGZ-Scalar theory in the non-minimal scheme is written as
\begin{equation}
Z^h_{\phi {\rm RGZ}} = \int [\EuScript{D}\mu]^h_{\phi\rm GZ}\,{\rm e}^{-S^{h,{\rm nm}}_{\phi {\rm RGZ}}+d\gamma^4 (N^2-1)V}\,,
\label{SubSect:NPBRST.10}
\end{equation}
with $[\EuScript{D}\mu]^h_{\phi\rm GZ} = [\EuScript{D}\mu]^h_{\rm GZ}[\EuScript{D}\phi]$. An important question to be addressed is: are the correlation functions of gauge-invariant operators computed with \eqref{SubSect:NPBRST.10} equivalent to those computed with the BRST-broken framework introduced in SubSect.~\ref{SubSect:NPS}? The answer is positive and the proof goes as follows: Consider the correlation function 
\begin{eqnarray}
&&\langle \EuScript{O}_1 (x_1)\ldots \EuScript{O}_n (x_n)\rangle^h_{\phi {\rm RGZ}} = \nonumber\\
&=& \frac{\int [\EuScript{D}\mu]^h_{\phi\rm GZ}\,\EuScript{O}_1 (x_1)\ldots \EuScript{O}_n (x_n)\,{\rm e}^{-S^{h,{\rm nm}}_{\phi {\rm RGZ}}+d\gamma^4 (N^2-1)V}}{\int [\EuScript{D}\mu]^h_{\phi\rm GZ}\,{\rm e}^{-S^{h,{\rm nm}}_{\phi {\rm RGZ}}+d\gamma^4 (N^2-1)V}}\,.\nonumber\\
\label{SubSect:NPBRST.11}
\end{eqnarray}
Next to that, let us integrate out the fields $b^a$, $\tau^a$ and $(\bar{\eta},\eta)^a$. This will produce the following term in the integrand,
\begin{equation}
\delta(\partial_\mu A^a_\mu)\delta(\partial_\mu (A^h)^a_\mu){\rm det}(-\partial_\mu D^{ab}_{\mu} (A^h))\,.
\label{SubSect:NPBRST.12}
\end{equation}
Using the functional generalization of the relation 
\begin{equation}
\delta (f(x)) = \frac{\delta (x-x_0)}{|f^\prime (x_0)|}\,,
\label{SubSect:NPBRST.13}
\end{equation}
with $x_0$ being the single root of the differentiable function $f(x)$ yields
\begin{equation}
\delta (\partial_\mu (A^h)^a_\mu) = \frac{\delta (\xi-\xi_0)}{{\rm det}(-\partial_\mu D^{ab}_{\mu} (A^h))}\,,
\label{SubSect:NPBRST.14}
\end{equation}
where $\xi_0 = \xi^a_0 T^a$ denotes the solution for $\xi$ of $\partial_\mu (A^h)_\mu = 0$. It reads
\begin{eqnarray}
\xi_0 &=& \frac{\partial A}{\partial^2} + \frac{ig}{\partial^2}\left[\partial A , \frac{\partial A}{\partial^2}\right]+\frac{ig}{\partial^2}\left[A_\mu,\partial_\mu  \frac{\partial A}{\partial^2}\right]\nonumber\\
&+&\frac{ig}{2}\frac{1}{\partial^2}\left[\frac{\partial A}{\partial^2},\partial A\right]+\EuScript{O}(A^3)\,,
\label{SubSect:NPBRST.15}
\end{eqnarray}
with $\partial A \equiv \partial_\mu A_\mu$. Every term of \eqref{SubSect:NPBRST.15} contains a divergence of the the gauge field. Due to the presence of the delta-functional imposing the Landau gauge condition in \eqref{SubSect:NPBRST.12}, one ends up with $\xi_0 = 0$. Hence, plugging Eqs.\eqref{SubSect:NPBRST.15} and \eqref{SubSect:NPBRST.14} into \eqref{SubSect:NPBRST.12} leads to
\begin{equation}
\delta (\partial_\mu A_\mu)\delta (\xi)\,.
\label{SubSect:NPBRST.16}
\end{equation}
Integrating over $\xi$, the delta-functional $\delta (\xi)$ decouples de Stueckelberg-like field and the dressed fields reduce to
\begin{equation}
(A^h)_\mu \to A_\mu \quad {\rm and} \quad \phi^h \to \phi\,.
\label{SubSect:NPBRST.17}
\end{equation}
Integrating out the localizing auxiliary fields, one gets $H(A^h)\to H(A)$ and $\mathcal{H}(\phi^h)\to \mathcal{H}(\phi)$. As a result, 
\begin{equation}
\langle \EuScript{O}_1 (x_1)\ldots \EuScript{O}_n (x_n)\rangle^h_{\phi {\rm RGZ}} = \langle \EuScript{O}_1 (x_1)\ldots \EuScript{O}_n (x_n)\rangle_{\phi {\rm RGZ}}\,.
\label{SubSect:NPBRST.18}
\end{equation}
Thus correlation functions of gauge-invariant operators are equivalent in the BRST-invariant and in the standard BRST-broken formulations of the RGZ-scalar action. In fact, the previous argument is more general and is applicable for correlation functions of the fields $A^a_\mu$ and $\phi^a$. As it is evident from \eqref{SubSect:NPBRST.16}, such an equivalence between correlation functions in different formulations of the RGZ-scalar partition function is a direct consequence of the transversality of the the dressed field $(A^h)_\mu$ together with the Landau gauge condition. This gives a special role to the Landau gauge since one can safely work with the simpler BRST-broken formulation in the computation of correlation functions of gluons and scalars. It is clear that by setting $\sigma^2 = 0$ the previous discussion remains untouched and therefore the equivalence remains valid in the minimal scheme.

\subsection{Gauge-invariant meaning of correlation functions in the Landau gauge\label{SubSect:GIMLG}}

The building blocks of observables in gauge theories are correlation functions of the elementary fields which in turn are not gauge-invariant. Yet the non-perturbative evaluation of such correlation functions became a meeting point of different approaches to non-perturbative YM and QCD over the last few decades. 

Nevertheless, the special role of the Landau gauge together with the dressing discussed in SubSect.~\ref{SubSec:BRSTRestSym} allow for a gauge-invariant meaning of correlation functions of elementary fields. This follows from the Eqs.\eqref{SubSect:NPBRST.17} and \eqref{SubSect:NPBRST.18}, i.e.,
\begin{eqnarray}
\langle A^{h,a_1}_{\mu_1} (x_1)\ldots A^{h,a_n}_{\mu_n} (x_n)\phi^{h,b_1} (y_1)\ldots \phi^{h,b_n} (y_n)\rangle^h_{\phi {\rm RGZ}}\nonumber\\
= \langle A^{a_1}_{\mu_1} (x_1)\ldots A^{a_n}_{\mu_n} (x_n) \phi^{b_1} (y_1)\ldots \phi^{b_n} (y_n)\rangle_{\phi {\rm RGZ}}\,.\nonumber\\
\label{SubSect:GIMLG.1}
\end{eqnarray}
The left-hand side of \eqref{SubSect:GIMLG.1} is manifestly gauge-invariant and thanks to the decoupling of the Stueckelberg-like field in the Landau gauge it equates to the correlation function of the elementary fields in such a gauge. In this work, we are concerned with the scalar field propagator in the Landau gauge. From \eqref{SubSect:GIMLG.1} it follows that
\begin{equation}
\langle \phi^{h,a} (x)\phi^{h,b} (y)\rangle^h_{\phi {\rm RGZ}}\
= \langle \phi^{a} (x) \phi^{b} (y)\rangle_{\phi {\rm RGZ}}\,.
\label{SubSect:GIMLG.2}
\end{equation}
As such, the computations here presented have a gauge-invariant meaning. Clearly, if computed in different gauges, the scalar-field propagator will differ but in every gauge, one can dress the scalar field and compute the correlation function in \eqref{SubSect:GIMLG.2} provided that the RGZ framework is consistently extended to such gauges. This result is valid both in the non-minimal and minimal schemes. From now on, we will narrow our focus to the minimal scheme. Our aim is to compute the one-loop correction to the scalar-field propagator in order to compare it with the available lattice data \cite{Maas:2018sqz}, and to make a few comments on the tree-level, standard one-loop corrected YM  and the non-minimal tree-level scheme results. 

\section{Scalar Field Propagator in the Minimal Scheme \label{Sec:SFPMS}}

In this section we provide a detailed account for the one-loop corrections to the scalar-field propagator in the RGZ-Scalar theory in the minimal scheme, i.e., given by Eq.\eqref{SubSect:MS.1}. Such a connected two-point function can be written as
\begin{equation}
\langle \phi^a (p) \phi^b (-p)\rangle = \delta^{ab}\mathcal{D}_{\phi} (p)\,.
\label{Sec:SFPMS.1}
\end{equation}
It can be extracted from the one-particle irreducible (1PI) functional $\Gamma$ through the relation\footnote{See Eq.~\eqref{Eq:conv.7} in Appendix~\ref{Sub:AP.Conv} for conventions.}
\begin{equation}
\sum_{\Phi_k} \Gamma^{(2)}_{\Phi_i \Phi_k} G_{\Phi_k \Phi_j} = \delta_{\Phi_i \Phi_j}\,.
\label{Sec:SFPMS.2}
\end{equation}
The notation $\Phi_i$ is employed to represent a generic field of RGZ-Scalar theory. By choosing $\Phi_i = \phi^a$ and $\Phi_j = \phi^b$ and taking the fields and sources to zero leads to
\begin{eqnarray}
\Gamma_{\phi^a \phi^c}^{(2)} G_{\phi^c \phi^b} &+& \Gamma^{(2)}_{\phi^a A^c_\mu} G_{A^c_\mu \phi^b} + \Gamma^{(2)}_{\phi^a b^c} G_{b^c \phi^b}\nonumber\\
&+& \Gamma^{(2)}_{\phi^a \varphi^{cd}_{\mu}} G_{\varphi^{cd}_{\mu} \phi^b} + \Gamma^{(2)}_{\phi^a \bar{\varphi}^{cd}_{\mu}} G_{\bar{\varphi}^{cd}_{\mu} \phi^b} = \delta^{ab}\,.\nonumber\\
\label{Sec:SFPMS.3}
\end{eqnarray}
From the Lautrup-Nakanishi field equation of motion, 
\begin{equation}
\frac{\delta \Gamma}{\delta b^a} = i\partial_\mu A^a_\mu,
\label{Sec:SFPMS.3.1}
\end{equation}
it follows that $\Gamma_{\phi^a b^c} = 0$. By acting with the test operator $\delta/\delta {J^b_{(\phi)} (y)}$ on \eqref{Sec:SFPMS.3.1} and taking sources and fields to zero leads to 
\begin{equation}
\partial^{x}_\mu\frac{\delta^2 W}{\delta J^a_{(A)\mu} (x) \delta J^{b}_{(\phi)}(y)} = \partial^x_\mu \langle A^a_{\mu} (x) \phi^b (y)\rangle = 0\,.
\label{Sec:SFPMS.4}
\end{equation}
In Fourier space it translates into
\begin{equation}
p_{\mu}\langle A^a_\mu (p) \phi^b (-p)\rangle \equiv p_{\mu} G_{A^a_{\mu}\phi^b} (p) = 0\,.
\label{Sec:SFPMS.5}
\end{equation}
Next to that we use color- and Lorentz-covariance to implement the following factorization of the tensor structures in \eqref{Sec:SFPMS.3},
\begin{eqnarray}
\Gamma^{(2)}_{\phi^a \phi^c} &=& \delta^{ac} \Gamma^{(2)}_{\phi \phi} \quad {\rm and} \quad G_{\phi^a \phi^c} = \delta^{ac}\mathcal{D}_\phi (p)\,,\nonumber\\
\Gamma^{(2)}_{\phi^a A^c_\mu} &=& \delta^{ac}p_\mu \Gamma^{(2)}_{\phi A} \quad {\rm and} \quad G_{\phi^a A^c_\mu} = \delta^{ac}p_\mu G_{\phi A}\,,\nonumber\\
\Gamma^{(2)}_{\phi^a {\varphi}^{cd}_{\mu}} &=& f^{acd}p_\mu \Gamma^{(2)}_{\phi \varphi} \quad {\rm and} \quad G_{\phi^a {\varphi}^{cd}_{\mu}} = f^{acd}p_\mu G_{\phi \varphi}\,,\nonumber\\
\Gamma^{(2)}_{\phi^a \bar{\varphi}^{cd}_{\mu}} &=& f^{acd}p_\mu \Gamma^{(2)}_{\phi \bar{\varphi}} \quad {\rm and} \quad G_{\phi^a \bar{\varphi}^{cd}_{\mu}} = f^{acd}p_\mu G_{\phi \bar{\varphi}}\,.\nonumber\\
\label{Sec:SFPMS.6}
\end{eqnarray}
Plugging Eq.\eqref{Sec:SFPMS.6} into \eqref{Sec:SFPMS.5} leads to $G_{A\phi} = 0$. Finally, Eq.\eqref{Sec:SFPMS.3} is reduced to
\begin{equation}
\Gamma^{(2)}_{\phi \phi} \mathcal{D}_{\phi} (p)+N\, \Gamma^{(2)}_{\phi \varphi} G_{\phi \varphi} + N\, \Gamma^{(2)}_{\phi \bar{\varphi}} G_{\phi \bar{\varphi}} = 1\,.
\label{Sec:SFPMS.7}
\end{equation}
At one-loop order, the last two terms of the left-hand side of Eq.\eqref{Sec:SFPMS.7} do not contribute. Thus, we can write
\begin{align}
	\mathcal{D}_{\phi} (p) = \left(\Gamma^{(2)}_{\phi \phi}\right)^{-1}.
\end{align}
Hence, our computation reduces to the evaluation of $\Gamma_{\phi\phi}$ at one-loop. 

In the RGZ framework, thanks to the restriction of the functional integral to the Gribov region, the tree-level gluon propagator is modified. It reads
\begin{align}
	\langle A_\mu^a(p) A_\nu^b(-p) \rangle_0 = \delta^{a b} \left(\delta_{\mu \nu} - \frac{p_\mu p_\nu}{p^2}\right) \mathcal{D}_A(p),
\end{align}
where we define the form factor $\mathcal{D}_A(p)$ as
\begin{equation}\label{eq:gluon.prop}
	\mathcal{D}_A(p) = \frac{p^2 + M^2}{\left(p^2 + M^2\right) \left(p^2 + m^2\right) + 2\gamma^4 g^2 N},
\end{equation}
and the parameters $(\gamma, m^2, M^2)$ were defined in Sect.~\ref{Sec:RGZOverview}. In order to rewrite the gluon propagator \eqref{eq:gluon.prop} closer to standard tree-level propagators, it is convenient to perform a partial fraction decomposition as follows,
\begin{align}
	\mathcal{D}_A(p) = \frac{R_{+}}{p^2 + \mu_{+}^2} + \frac{R_{-}}{p^2 + \mu_{-}^2},
\end{align}
where
\begin{align}
	R_{\pm} = \frac{1}{2} \left(1 \pm \frac{m^2 - M^2}{\Omega}\right) , \quad \mu_{\pm}^2 = \frac{1}{2} \left(m^2 + M^2 \pm \Omega\right),
\end{align}
with $\Omega = \sqrt{\left(m^2 - M^2\right)^2 - 8\gamma^4 g^2 N}$.

The RGZ gluon propagator Eq.~\eqref{eq:gluon.prop} is suppressed in the infrared, and {is compatible with a non-vanishing value at zero momentum, in accordance with lattice results. 
In fact, using this tree-level RGZ gluon propagator, one can fit the parameters $M$, $m$, and $\gamma$ by comparing it with lattice data. Notably, the fitting based on $SU(2)$ and $SU(3)$ lattice data shows that the RGZ tree-level gluon propagator is sufficient to reproduce the lattice data in the infrared~\cite{Cucchieri:2011ig, Oliveira:2012eh}.
We emphasize that, although being determined by a fitting procedure with lattice data, the parameters $M$, $m$, and $\gamma$ are not free parameters, and can in principle be determined self-consistently by solving their corresponding gap equations~\cite{Dudal:2019ing}.

As previously discussed, at one-loop, in order to compute the scalar-field propagator, it is just necessary to evaluate the inverse of the 1PI scalar two-point function. There are three diagrams that contribute to $\Gamma_{\phi\phi}^{(2)}$ at one-loop (c.f. Fig.~\ref{fig:diagrams}): a scalar tadpole ($\ScalarTad$), a gluon tadpole ($\GluonTad$) and a mixed sunset ($\MixedSun$).
To compute the diagrams in Fig. \ref{fig:diagrams}, we used the \texttt{Mathematica} packages \textit{xAct} \cite{xAct1,xAct2}, \textit{Form-Tracer} \cite{FormTracer} and \textit{FeynCalc} \cite{FeynCalc1,FeynCalc2,FeynCalc3}.

\begin{figure}
	\includegraphics[width=\linewidth]{Diagrams_1Loop.pdf}
	\caption{Diagrams contributing to the one-loop 1PI scalar two-point function $\Gamma^{(2)}_{\phi\phi}$. From left to right, they correspond to $\ScalarTad$, $\GluonTad$ and $\MixedSun$}
	\label{fig:diagrams}
\end{figure}

Therefore, we have for the 1PI scalar two-point function,
\begin{align}\label{eq:2pointScalar}
	\Gamma^{(2)}_{\phi\phi} = \left(p^2 + m_\phi^2\right) + \left(\ScalarTad + \GluonTad + \MixedSun \right) +  \CounterTerm \,,
\end{align}
where $\CounterTerm$ denotes the counter-terms that should be chosen to cancel the divergences and to ensure the renormalization conditions. In the present case, the ultraviolet divergences are either momentum independent, or proportional to $p^2$, thus $\CounterTerm = \delta_\phi \, p^2 + \delta_{m_\phi}$. 

Computing the diagrams in Fig.~\ref{fig:diagrams} within dimensional regularization yields
\begin{align}
	\ScalarTad &= - \frac{\lambda \,m_{\phi}^2 (N^2 + 1) }{96 \pi^2} \frac{1}{\varepsilon} + \ScalarTadFin \,, \\
	\GluonTad &= - \frac{3 N g^2  \left(\mu_{+}^2 R_{+} + \mu_{-}^2 R_{-}\right)}{16 \pi^2} \frac{1}{\varepsilon} + \GluonTadFin \,, \\
	\MixedSun &= - \frac{3 p^2 N g^2  \left(R_{+} + R_{-}\right) }{16 \pi^2} \frac{1}{\varepsilon} + \MixedSunFin \,.
\end{align}
The finite parts are given by
\begin{align}
	\ScalarTadFin = \frac{\lambda m_{\phi}^2 (N^2 + 1) }{96 \pi^2} \left[ \log\left(\frac{m^2_\phi}{4 \pi \mu^2}\right) + \gamma -1 \right] \,,
\end{align}
for the scalar tadpole,
\begin{align}
	\GluonTadFin = \frac{ 3 N g^2  }{16 \pi^2} &\left[ \mu_{+}^2 R_{+} \log \left( \frac{\mu_{+}^2}{4 \pi \mu^2}\right) + \mu_{-}^2 R_{-} \log \left( \frac{\mu_{-}^2}{4 \pi \mu^2}\right) \right. \nonumber \\
	&\left.  + \left( \gamma -\frac{1}{3}\right) \left(\mu_{+}^2 R_{+} + \mu_{-}^2 R_{-} \right)   \right] \,,
\end{align}
for the gluon tadpole, and
\begin{align}
	&\MixedSunFin = \frac{N g^2 p^2}{4 \pi^2} \left[ \frac{3}{4} (R_{+} + R_{-}) \left(\gamma - \log 4 \pi\right) \right. \nonumber \\
	&\left. \quad +\int_0^1  \mathrm{d}x \left(R_{+} \log \frac{\Delta_{+}}{\mu^2} + R_{-} \log \frac{\Delta_{-}}{\mu^2}\right) \right. \nonumber \\
	&\left. \quad+    \int_0^1  \mathrm{d}x  \int_0^{1-x}  \mathrm{d}y \, p^2 x^2 \left(\frac{R_{+}}{\tilde{\Delta}_{+}} + \frac{R_{-}}{\tilde{\Delta}_{-}} \right)    \right. \nonumber \\
	&\left. \quad-\frac{1}{2} \int_0^1 \mathrm{d}x \int_0^{1-x} \mathrm{d}y \left(R_{+} \log \frac{\tilde{\Delta}_{+}}{\mu^2} + R_{-} \log \frac{\tilde{\Delta}_{-}}{\mu^2} \right)  \right] \,,
\end{align}
for the mixed sunset. In the last expression, we used the shorthand notation $\Delta_i = p^2 \,x (1-x) + (m_\phi^2 - \mu_i^2)x + \mu_i^2$ and $\tilde{\Delta}_i = p^2x (1-x) + m_\phi^2x + \mu_i^2 y$. We point out that the only momentum-dependent contribution arises in the mixed sunset $\MixedSun$diagram.

The simplest scheme to renormalize $\Gamma^{(2)}_{\phi\phi}$ is the minimal-subtraction-bar scheme ($\overline{\text{MS}}$). In this case, the renormalized two-point function reads
\begin{align}\label{key}
	\Gamma^{(2)}_{\phi\phi, \overline{\text{MS}}} &= p^2 + m_\phi^2 + \frac{\lambda\, m_{\phi}^2 (N^2 + 1) }{96 \pi^2} \left[ \log\left(\frac{m^2_\phi}{ \mu^2}\right)  -1 \right] \nonumber \\
	&+ \frac{ 3 N g^2  }{16 \pi^2} \left[ \mu_{+}^2 R_{+} \log \left( \frac{\mu_{+}^2}{ \mu^2}\right) + \mu_{-}^2 R_{-} \log \left( \frac{\mu_{-}^2}{ \mu^2}\right) \right. \nonumber \\
	&\left.   -\frac{1}{3} \left(\mu_{+}^2 R_{+} + \mu_{-}^2 R_{-} \right)   \right] \nonumber \\
	&+ \frac{N g^2 p^2}{4 \pi^2} \left[ \int_0^1  \mathrm{d}x \left(R_{+} \log \frac{\Delta_{+}}{\mu^2} + R_{-} \log \frac{\Delta_{-}}{\mu^2}\right) \right. \nonumber \\
	&\left.  -\frac{1}{2} \int_0^1 \!\! \mathrm{d}x \int_0^{1-x} \!\!\!\!\!\! \mathrm{d}y \left(R_{+} \log \frac{\tilde{\Delta}_{+}}{\mu^2} + R_{-} \log \frac{\tilde{\Delta}_{-}}{\mu^2} \right)    \right. \nonumber \\
	&\left.  +    \int_0^1 \! \mathrm{d}x  \int_0^{1-x}\!\!\!  \mathrm{d}y \, p^2 x^2 \left(\frac{R_{+}}{\tilde{\Delta}_{+}} + \frac{R_{-}}{\tilde{\Delta}_{-}} \right)  \right].
\end{align}
However, as we are interested in comparing our results with lattice data, it is more appropriate to work with a renormalization scheme similar to what is usually employed in lattice calculations. 
Thus, we impose the following renormalization conditions
\begin{eqnarray}
	\frac{\mathrm{d}}{\mathrm{d} p^2} \Gamma^{(2)}_{\phi\phi}\vert_{p=\mu} &=& 1  \label{MOM1} \\
	\Gamma^{(2)}_{\phi\phi}\left(p=\mu\right) &=& \mu^2 + m_\phi^2 \,, \label{MOM2}
\end{eqnarray}
defining the momentum subtraction scheme. 
In this case, the resulting renormalized expression is too lengthy to be reported here. The full expression is available in an ancillary \texttt{Mathematica} notebook.

\section{Comparing the propagator in the minimal scheme with Lattice Data \label{Sec:CompLattice}}
\begin{figure*}[t!]
	\includegraphics[width=\linewidth]{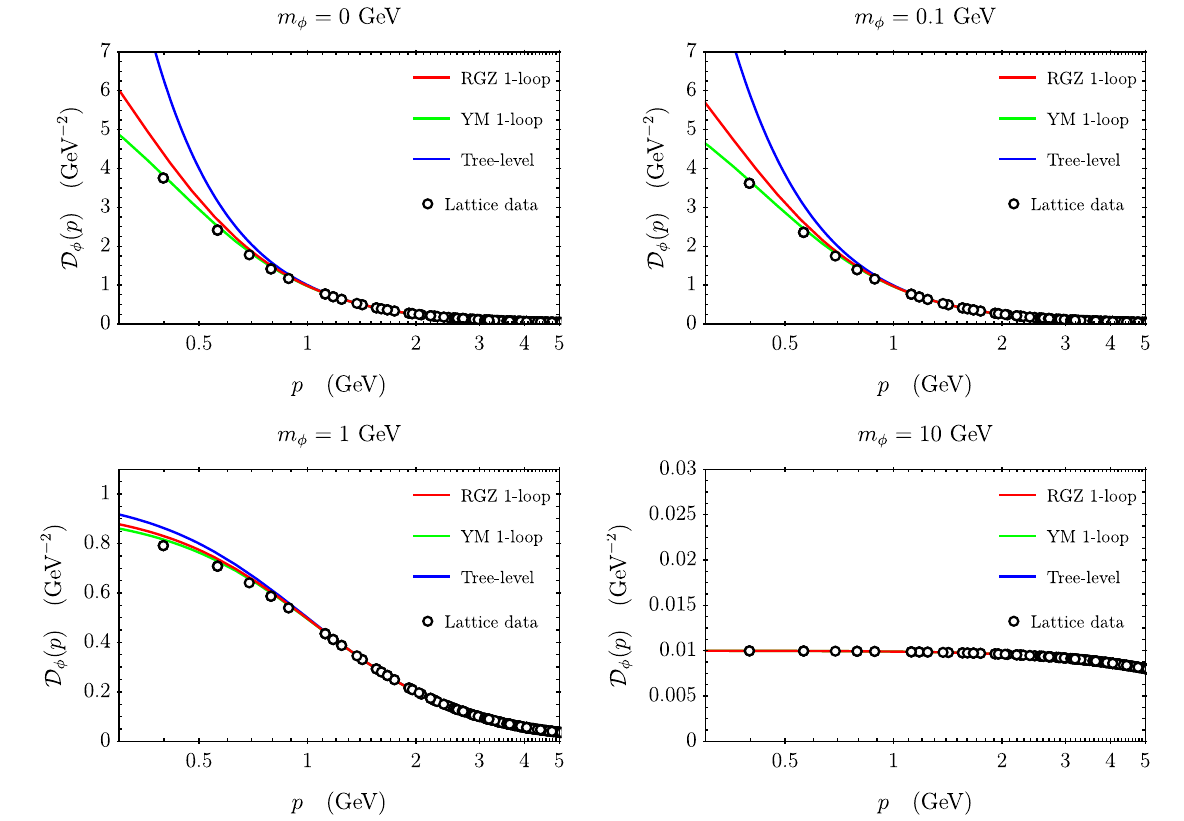}
	\caption{Renormalized scalar propagators in the adjoint representation at one-loop (red line) in units of $\text{GeV}^{-2}$, for masses $m_\phi = 0, 0.1, 1$, and $10$ GeV. We compare the RGZ results at one-loop (red line) with lattice data (white dots), the tree-level result (blue line) and the standard YM result at one-loop (green line).}
	\label{ScalarProps}
\end{figure*}

\begin{figure*}[t!]
	\includegraphics[width=\linewidth]{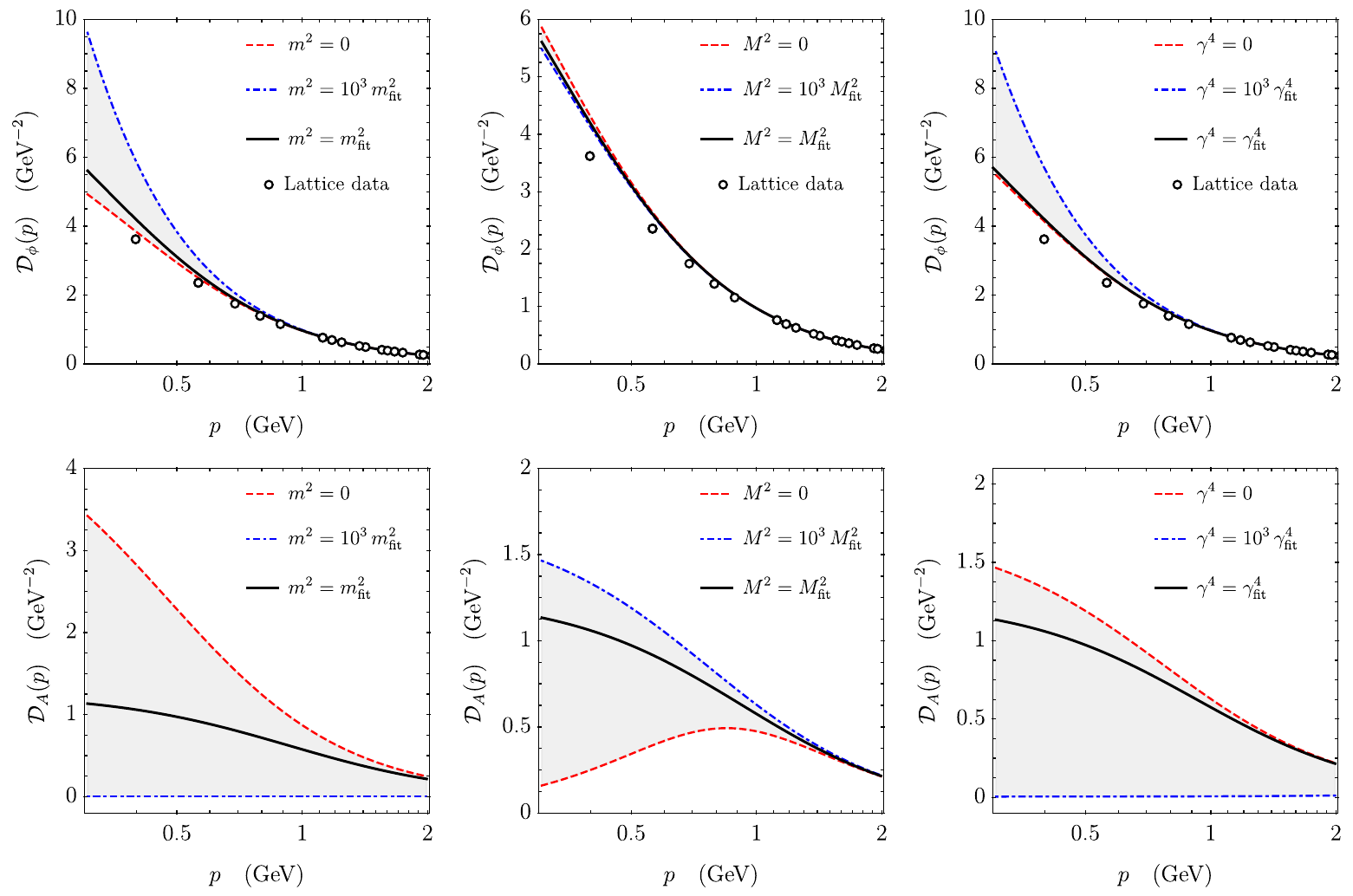}
	\caption{We study the effect of variations of Gribov parameters on the scalar and tree-level gluon propagators. In the upper panel, we show the scalar propagator. In the lower panel, we show the tree-level RGZ gluon propagator.
		In the first column, we study variations of the mass parameter $m^2$ (with $M^2 = M^2_\textmd{fit}$ and $\gamma^4 = \gamma^4_\textmd{fit}$). In the second column, we study variations of the mass parameter $M^2$ (with $m^2 = m^2_\textmd{fit}$ and $\gamma^4 = \gamma^4_\textmd{fit}$). In the third column, we study variations of the mass parameter $\gamma^4$ (with $m^2 = m^2_\textmd{fit}$ and $M^2 = M^2_\textmd{fit}$). In all plots, we set the scalar mass $m_\phi=0.1\, \textmd{GeV}$. We point out that the horizontal axis starts at $p=0.3 \,\text{GeV}$.}
	\label{ScalarPropsVaryingRGZ}
\end{figure*}

\begin{figure}[t!]
	\includegraphics[width=\linewidth]{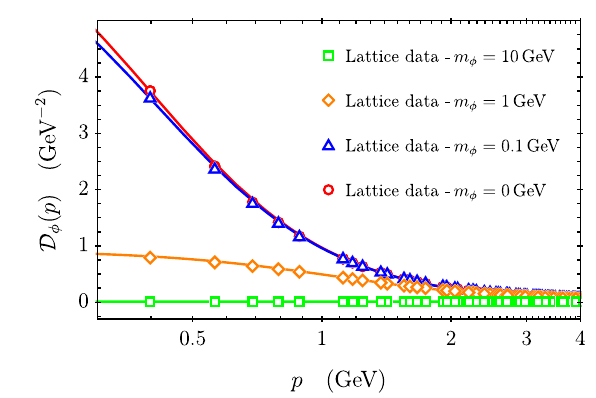}
	\caption{Scalar propagator at one-loop with gauge coupling $g^2 = 2.52$. We use Gribov parameters determined by the fitting of the tree-level gluon propagator (c.f.~\eqref{eq:GribovFit}).
		We use the lattice corresponding to gauge coupling $g^2 = 1.621$ \cite{Maas:2018sqz}}
	\label{FitSimult}
\end{figure}

In this section, we compare our results for the one-loop scalar propagator $\mathcal{D}_\phi(p)$ with lattice data reported in \cite{Maas:2018sqz}.
For a consistent comparison, we need to renormalize the 1PI two-point function for the scalar field at one-loop order reported in the last section in a way compatible with schemes employed in the lattice. Here we use the lattice data from ~\cite{Maas:2018sqz} and for this reason we adopt the momentum subtraction scheme defined in Eqs.~\eqref{MOM1} and \eqref{MOM2}.

The lattice data that we use for the purposes of the present comparison were taken considering lattice volumes $V= (32)^4$. Analyzing different lattice cutoffs, we observed a similar qualitative behavior in the renormalized scalar propagators. Thus we decided to present only the results for one choice of lattice cutoff, $a^{-1} = 2.03 \, {\rm GeV}$. Associated with this data, we have $\beta = 4/g^2 = 2.467$, corresponding to $g^2 = 1.621$. In what follows, we consider the renormalized scalar propagators in the adjoint representation of $SU(2)$ gauge group, with four different values of the renormalized scalar mass: $m_\phi = 0\,\textmd{GeV}$, $m_\phi =  0.1\,\textmd{GeV}$,  $m_\phi =  1.0\,\textmd{GeV}$, and $m_\phi =  10 \,\textmd{GeV}$. We work with renormalization scale $\mu = 1.5 \, \text{GeV}$. 

Concerning the mass parameters present in the RGZ scenario, i.e., $\left( \gamma, m^2, M^2 \right)$, we use the values obtained by fitting the tree-level RGZ gluon propagator (c.f., Eq.~\eqref{eq:gluon.prop}) with lattice data \cite{Cucchieri:2011ig}:
\begin{equation}\label{eq:GribovFit}
	\begin{aligned}
		m^2_\textmd{fit} &= 0.768^2 \,\textmd{GeV}^2 \,,\\
		M^2_\textmd{fit} &= 2.508 \,\textmd{GeV}^2 \,,\\
		\gamma^4_\textmd{fit} &= (0.360/g)^2\,\textmd{GeV}^4  \,.
	\end{aligned}
\end{equation}
Let us stress that, as it will be shown later on, a considerable variation of the mass parameters arising from the gluon propagator in the RGZ setup will not affect significantly the scalar-field propagator. Yet due to the well-known successful fitting of the tree-level gluon propagator with available lattice data this would entail a drastic change in the gluonic sector. Hence, we take the values of the mass parameters obtained from a particular fitting of the tree-level gluon propagator with the lattice data just as a benchmark.

In Fig.~\ref{ScalarProps}, we exhibit the renormalized scalar propagator at one-loop order in the minimal scheme, for four different choices of the scalar mass parameter $m_\phi$. We show the lattice data, the tree-level scalar propagator, the one-loop scalar propagator in the minimal scheme, and the one-loop result for the scalar propagator in the standard YM setting. 
For each value of the scalar mass $m_\phi$, the lattice data points and the three other curves are collected in a single plot for the sake of comparison. In the regime of high momentum ($p \gtrsim 1 \, {\rm GeV}$), we can see that all the three curves depicted in Fig.~\ref{ScalarProps} are converge to the lattice data.

In the infrared region ($p \lesssim 1 \, {\rm GeV}$), we observe differences between the curves shown in Fig.~\ref{ScalarProps}. First, we note that the infrared behavior of the tree-level propagator exhibits a significant difference in comparison with the lattice results. Second, the one-loop correction in the minimal scheme improves the infrared behavior of $\mathcal{D}_\phi(p)$. However it is not sufficient to reproduce the lattice data in the deep infrared. Third, and perhaps more surprising, the one-loop result in the standard YM framework reproduces the lattice data in the infrared regime satisfactorily. 
We note that the difference between the results is suppressed by the scalar mass $m_\phi$, such that for $m_\phi = 10 \,\textmd{GeV}$ all the curves (including the tree level result) reproduce the lattice data quite well. 

Two comments are in order: The one-loop scalar propagator computed in the standard YM setting provides very satisfactory results in the infrared. The result is indeed a prediction since no fittings were necessary in this case. However, the theory must be self-consistent and different correlation functions should also be well described by a particular choice of setup. In this case, YM theories coupled to scalars do not have a gluon propagator that reproduces well lattice data in this perturbative scheme and therefore just being able to properly describe the scalar sector does not seem to be a consistent choice. It could be that a non-perturbative treatment for the computation of the gluon propagator is compatible with the lattice data and that the relevant features of the scalar sector are properly captured in perturbation theory. We are not able to make a statement in this regard in the present work. We refer the reader to works on Scalar-YM systems within the formalism of Dyson-Schwinger equations~\cite{Fister:2010yw,Macher:2011ys}. Second, the one-loop scalar propagator in the minimal scheme involves the use of fittings for the mass parameters which in turn were extracted from different lattice data. The systematic error carried over different simulations could very well entail a modification on those values which would affect the shape of form factor. It is beyond our capabilities to perform a systematic investigation in this regard and conclude whether the results could be improved. Instead, we employ a cheaper strategy which is a parametric analysis of the impact of the values of those mass parameters to the gluon and scalar propagators.

In the following, we study the impact of variations of the mass parameters $(\gamma,m^2,M^2)$ on the results of the scalar propagator $\mathcal{D}_\phi(p)$. To estimate the impact of each one of the mass parameters, we perform three independent analysis, in each case we treat one of the parameters as free, while fixing the other two parameters by their fitted values (c.f. Eq.~\eqref{eq:GribovFit}). In Fig.~\ref{ScalarPropsVaryingRGZ}, we show the main results concerning the variation of the mass parameters, and their impact on the scalar propagator. To complement the analysis, in Fig.~\ref{ScalarPropsVaryingRGZ} (lower panel), we also show how the tree-level RGZ gluon propagator is affected by variations of the mass parameters. 
In the first column, we study variations of the mass parameter $m^2$ (while setting $M^2 = M^2_\textmd{fit}$ and $\gamma^4 = \gamma^4_\textmd{fit}$). In this case, we observe that reducing the value of $m^2$ seems to improve the infrared behavior of $\mathcal{D}_\phi(p)$, while larger values of $m^2$ push $\mathcal{D}_\phi(p)$ further away from the lattice data. 
In the second column, we study variations of the mass parameter $M^2$ (while setting $m^2 = m^2_\textmd{fit}$ and $\gamma^4 = \gamma^4_\textmd{fit}$). In this case, we observe that larger values of $M^2$ push $\mathcal{D}_\phi(p)$ slightly closer to the lattice data, which smaller values of $M^2$ push $\mathcal{D}_\phi(p)$ further away from the lattice data. Notably, $\mathcal{D}_\phi(p)$ seems to be less sensitive to variations of $M^2$ in comparison with the other mass parameters.
In the third column, we study variation of the Gribov parameter $\gamma$ (while setting $m^2 = m^2_\textmd{fit}$ and $M^2 = M^2_\textmd{fit}$). In this case, we note that reducing $\gamma$ has almost no effect on $\mathcal{D}_\phi(p)$, while enlarging of $\gamma$ pushes $\mathcal{D}_\phi(p)$ further away from the lattice data.

It is important to emphasize that in all cases, the tree-level RGZ gluon propagator seems to be very sensitive to variations of the mass parameters (c.f.~Fig.~\ref{ScalarPropsVaryingRGZ}, lower panel). Thus our analysis suggests that there is little room for the improvement of infrared regime of $\mathcal{D}_\phi(p)$ based only on variations of the parameters $(\gamma,m^2,M^2)$. As we can see in Fig.~\ref{ScalarPropsVaryingRGZ}, small changes in the scalar-propagator $\mathcal{D}_\phi(p)$ translate into significant changes in the gluon propagator $\mathcal{D}_A(p)$. Of course, this conclusion comes with the caveats that the gluon propagator is the tree-level one and that we are probing variations of the mass parameters independently but as stated before, they are not free and should satisfy gap equations.

So far we have focused on results obtained with gauge coupling fixed by $g^2 = 1.621$, which is the same value adopted in the lattice simulations that generated the data reported in \cite{Maas:2018sqz}.
Aiming at an improvement of our results obtained in the minimal scheme, we also considered an alternative approach, where we treat the gauge coupling as a fitting parameter (while using \eqref{eq:GribovFit}). In this case, we have found that the choice $g^2 = 2.52$ results in a significant improvement of $\mathcal{D}_\phi(p)$ in the infrared regime, allowing us to reproduce the lattice data for all values of $m_\phi$ under consideration. In Fig.~\ref{FitSimult}, we show the results corresponding to $g^2 = 2.52$.

It is important to remark that, even though our results do not perfectly match the lattice data for the deep infrared region, there is still room for improvement. First of all, we adopted a constant gauge coupling for all momenta, fixed in the lattice simulations at the renormalization scale $\mu = 1.5 \, \text{GeV}$, but we know that there is a considerable running of the gauge coupling in the infrared. 
Considering that we were able to fit the lattice data by changing the value of the gauge coupling, it is reasonable to expect that we would be able to find a better agreement with lattice data by considering the running of the gauge coupling. However, performing a renormalization group improvement in the RGZ scenario would require a more involved computation, that is beyond the scope of the present work.  
Furthermore, we expect to find a better agreement with the lattice data by fixing the Gribov parameters using a fitting of the gluon propagator in the RGZ scenario at one-loop, instead of the tree-level fitting values that we have used here. The one-loop computation of the gluon propagator in the RGZ scenario and its comparison with the lattice data is work in progress, and it will be reported in another paper. 

\section{Remarks on positivity violation  \label{Sec:Positivity}}

\begin{figure*}[t!]
	\includegraphics[width=\linewidth]{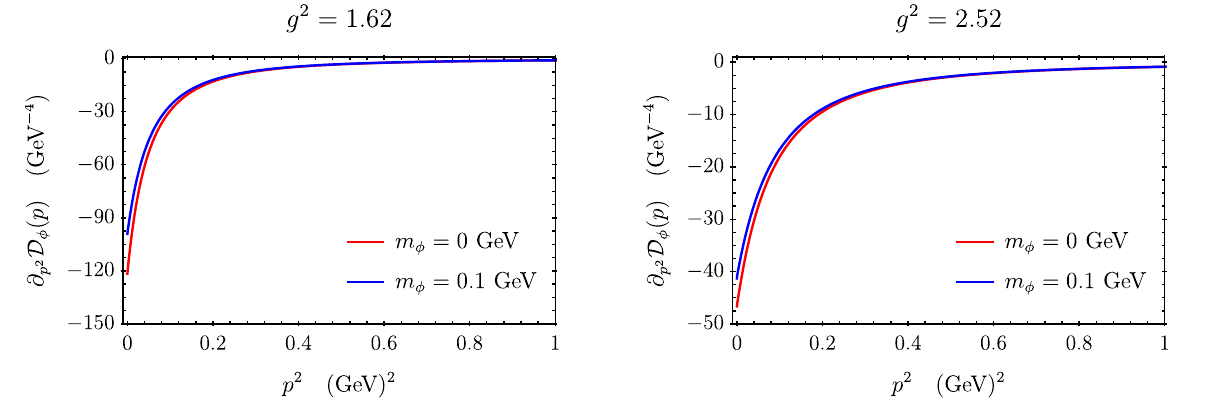}
	\caption{We show $\partial_{p^2} \mathcal{D}_\phi(p)$ as a function of $p^2$ for two choices of the gauge coupling: $g^2 = 1.62$ (left panel) and $g^2 = 2.52$ (right panel). In both cases, we fixed the mass parameters as $(\gamma^4_\textmd{fit},m^2_\textmd{fit},M^2_\textmd{fit})$ (c.f., Eq.~\eqref{eq:GribovFit}).}
	\label{Fig:Positivity}
\end{figure*}

\begin{figure*}[t!]
	\includegraphics[width=\linewidth]{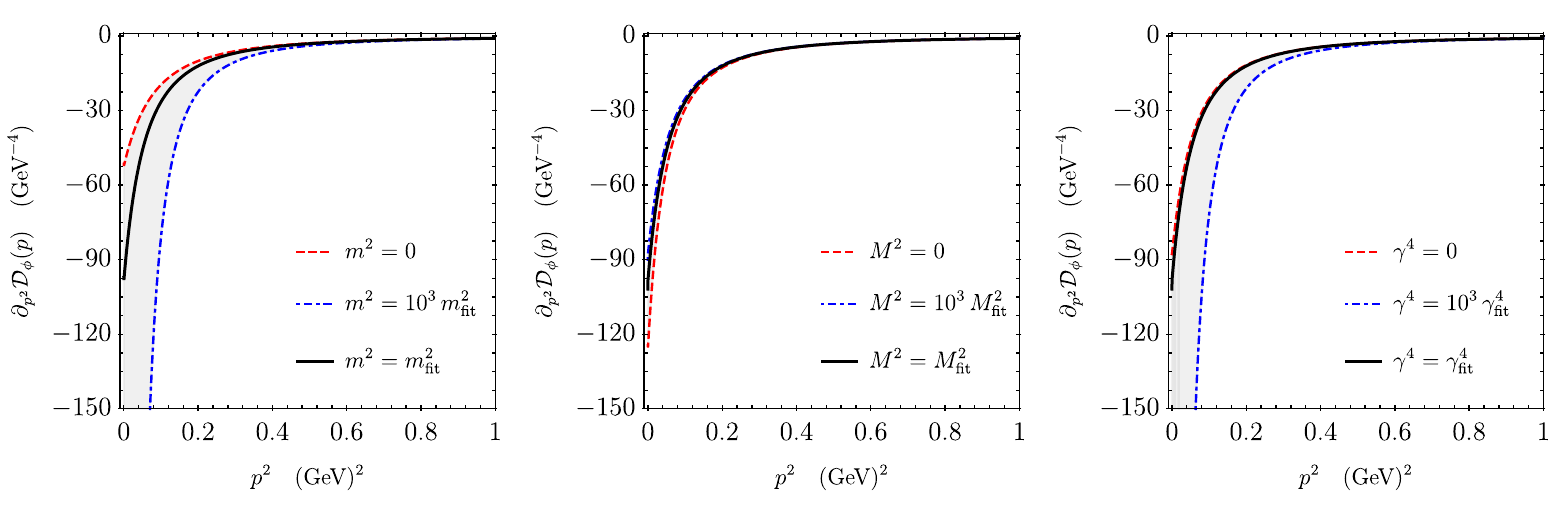}
	\caption{We study the effect of variations of the mass parameters on $\partial_{p^2} \mathcal{D}_\phi(p)$.
	In the first column, we study variations of the mass parameter $m^2$ (with $M^2 = M^2_\textmd{fit}$ and $\gamma^4 = \gamma^4_\textmd{fit}$). In the second column, we study variations of the mass parameter $M^2$ (with $m^2 = m^2_\textmd{fit}$ and $\gamma^4 = \gamma^4_\textmd{fit}$). In the third column, we study variations of the mass parameter $\gamma^4$ (with $m^2 = m^2_\textmd{fit}$ and $M^2 = M^2_\textmd{fit}$). In all plots, we set scalar mass $m_\phi=0.1\, \textmd{GeV}$}
	\label{Fig:Positivity_Parameterized}
\end{figure*}

In this section we search for signs of positivity violation based on our results for the scalar propagator $\mathcal{D}_\phi(p)$. In particular, we test if the derivative of $\mathcal{D}_\phi(p)$ with respect to $p^2$ (\textmd{i.e.}, $\partial_{p^2} \mathcal{D}_\phi(p)$) has zeros. A zero of $\partial_{p^2} \mathcal{D}_\phi(p)$ translates into an extremum for the propagator $\mathcal{D}_\phi(p)$, hence, indicating positivity violation. Being a colored matter field, the scalars should be confined as the gluons. Positivity violation of the propagator indicates that potentially one cannot interpret the associated excitations as part of the physical spectrum. This discussion, however, is rather subtle. To our knowledge, there is no direct association between confinement and positivity violation. Yet it is well-established from lattice simulations that the gluon propagator violates reflection positivity and this is interpreted as a trace of confinement. Nevertheless, the propagators are gauge-dependent quantities and so is its spectral representation. Hence it becomes much less clear how to assign a physical interpretation to such a violation. Tentatively, the two-point function of the dressed scalar fields as defined in Eq.~\eqref{SubSect:NPSBRST.1} could provide a gauge-invariant object that captures positivity violation. However, this deserves more investigation. In any case, we find important to report our analysis of this matter for the scalar sector.

The derivative test is a sufficient  but not necessary condition to verify positivity violation. The reasoning goes as follows: Consider the K\"all\'en-Lehmann representation of the propagator $\mathcal{D}(p^2)$, i.e.,
\begin{equation}
\mathcal{D}(p^2) = \int^{\infty}_0 {\rm d}\mu\frac{\rho (\mu)}{p^2+\mu}\,,
\label{Eq:PosViol.1}
\end{equation}
with the spectral density being denoted by $\rho (\mu)$. If the spectral density is positive then
\begin{equation}
\frac{\partial\mathcal{D}(p^2)}{\partial p^2} = -\int^{\infty}_0 {\rm d}\mu\frac{\rho (\mu)}{(p^2+\mu)^2}\,,
\label{Eq:PosViol.2}
\end{equation} 
is a strictly negative function of $p^2$. If $\partial_{p^2}\mathcal{D}(p^2)$ has a root, this entails positivity violation of $\rho (\mu)$. Hence, searching for zeros of $\partial_{p^2}\mathcal{D}(p^2)$ is a sufficient condition for positivity violation. However, it should be clear that it is not necessary. Moreover, there is an extra assumption in this test: we are assuming that the propagator $\mathcal{D}(p^2)$ has a spectral representation. Clearly, one can keep taking derivatives which will lead to alternating signs in front of the integral. Thus, for a positive spectral function, the derivatives of the form factor have well defined signs and any zero will signal positivity violation.

In Fig. \ref{Fig:Positivity}, we plot $\partial_{p^2} \mathcal{D}_\phi(p)$ as a function of $p^2$ for two different choices of gauge couplings ($g^2 = 1.62$ and $g^2=2.52$), and setting the mass parameters to their fitted values $(\gamma^4_\textmd{fit},m^2_\textmd{fit},M^2_\textmd{fit})$ (c.f.~Eq.~\eqref{eq:GribovFit}). We focus on the results obtained with scalar masses $m_\phi=0 \,\text{GeV}$ and $m_\phi=0.1 \,\textmd{GeV}$, but we have verified that the choices $m_\phi=1.0 \,\textmd{GeV}$ and $m_\phi=10 \,\textmd{GeV}$ lead to qualitatively similar results. We restrict the analysis to the range $p^2\leq1 \,\textmd{GeV}$, since we are looking for positivity violation in the infrared regime.
As we see in Fig. \ref{Fig:Positivity}, $\partial_{p^2} \mathcal{D}_\phi(p)$ does not have any zeros, making impossible to establish any conclusions regarding positivity violation from this analysis.

To complement the analysis, we also studied the impact of variations of the mass parameters on $\partial_{p^2} \mathcal{D}_\phi(p)$. We show the main results of such analysis in Fig. \ref{Fig:Positivity_Parameterized}. As we see, variations on the Gribov parameters do not change our conclusions concerning the non-existence of zeros of $\partial_{p^2} \mathcal{D}_\phi(p)$. Hence, we cannot draw any conclusions regarding positivity violation from this parametric analysis and take the definite sign of $\partial_{p^2} \mathcal{D}_\phi(p)$ robust against changes on the mass parameters.

As a final remark, we have also investigated positivity violation from the perspective of the Schwinger function, defined as
\begin{align}\label{eq:SchwingerFunct}
	\mathcal{C}_\phi(t) = \frac{1}{2\pi} \int_{-\infty}^{\infty} {\rm d}p \, \mathcal{D}_\phi(p) {\rm e}^{-ipt} \,.
\end{align}
A positive spectral function implies that $\mathcal{C}_\phi(t)$ should be positive for all values of $t$. In this sense, negative values of $\mathcal{C}_\phi(t)$ provides hints towards positivity violation. In our analysis based on the Schwinger function, we found no indications for positivity violation for small values of $t$. For large values of $t$, the results are contaminated by numerical instabilities due to rapid oscillations of the integrand in Eq. \eqref{eq:SchwingerFunct}. Thus, our searches for positivity violation based on the Schwinger function were also inconclusive. Therefore, the present analysis does not provide any evidence for positivity violation in the scalar sector. Yet the tests that we explored are all sufficient but not necessary for positivity violation and hence this topic deserves further explorations that we leave for future work.

\section{Comments on the non-minimal scheme \label{Sec:CompMSNPS}}

The non-minimal scheme provides an interesting and reasonable proposal for coupling matter fields to the RGZ framework. In such a scheme the matter fields feel the the restriction of the path integral measure to the Gribov horizon already at the tree-level. 
This is appealing at a first sight because the RGZ-like form of the propagator for matter fields seems to be sufficient to fit the available lattice data as reported in \cite{Capri:2014bsa}. However, the obvious drawback of this approach is that one has two new mass parameters that can be adjusted to fit the data irrespective of the data arising from the gluon sector. Hence, it becomes clear that it would be desirable to derive the properties of the matter correlation functions without the need of introducing such an extra structure that does not share the same geometrical justification as for the gauge-field sector. It turns out that our results indicate that the non-minimal scenario does not seem to be necessary 
to reproduce the available lattice data in the scalar sector. In fact, the one-loop computation performed in the minimal scheme already gives a fairly good agreement with lattice data, without the need of any extra assumption compared to the RGZ scenario for pure YM theories.

Yet we must emphasize that this conclusion is taken just analyzing the scalar propagator. A clear direction that is under investigation is what happens in the more physically relevant sector of quarks. Our tentative justification for the sufficiency of the minimal scheme relies on the  recent analysis of the Curci-Ferrari model coupled to quarks see, e.g., \cite{Pelaez:2017bhh}. In this model, the lattice data for the gluon propagator is well described at one-loop order while the quark sector needs higher-order loop corrections to achieve quantitative precision. The analogy in the present case is that the tree-level RGZ gluon propagator describes well the lattice data while the matter sector requires loop correction to allow for the non-perturbative information of the gluon sector to manifest itself in the loops. Yet it is interesting to explore the non-minimal scheme as a potential candidate to effectively describe the outcome of the inclusion of higher loops in the matter sector. This is an open problem that is left for future investigation.

As a final remark regarding the tree-level scalar propagator in the non-minimal scheme, since it has exactly the same structure of the RGZ propagator for the gluon and this is known to violate reflection positivity, one might wonder about the status of positivity violation in the scalar sector in such a scheme. Taking the derivative with respect to $p^2$ of Eq.~\eqref{SubSect:NPS.7} (and neglecting the color structure) leads to
\begin{equation}
\partial_{p^2} \mathcal{D}^{\rm nm}_\phi (p^2) = \frac{2Ng^2\sigma^4 - (p^2+M^2_\phi)^2}{[(p^2+m^2_\phi)(p^2+M^2_\phi)+2Ng^2\sigma^4]^2}\,.
\label{Eq:CNMS.1}
\end{equation}
The expression that gives the zeros of Eq.~\eqref{Eq:CNMS.1} is a quadratic polynomial on $p^2$ and it is clear that positive real roots are viable but not necessary. Just with the fitted values for the mass parameters one could tell if positivity is violated through the derivative test. Remarkably if one takes the derivative of Eq.~\eqref{Eq:CNMS.1} and searches for roots, the polynomial is cubic on $p^2$ and therefore it must have a real root. What remains to be verified is whether such a real root is positive or not. In the case of being positive, reflection positivity is violated. Taking the fitted values from \cite{Capri:2014bsa} in the case of $m_\phi = 0 \,\textmd{GeV}$ we find no evidence for positivity violation using the derivative tests, a fact that qualitatively agrees with the minimal-scheme results.

\section{Conclusions \label{Sec:Conclusions}}

The RGZ framework provides a local and renormalizable setup to eliminate infinitesimal Gribov copies and also accounting for the dynamical generation of condensates. It provides an improvement with respect to the standard FP action in the sense that it takes seriously the non-uniqueness of the selection of gauge configurations along a gauge orbit. The effects driven by the removal of such spurious configurations become relevant in the infrared where the theory is strongly coupled.

At the same time, the inclusion of matter fields in this improved gauge-fixing procedure brings up several important questions such as: Should we modify the standard prescription of coupling matter to gauge fields? How the dynamics of colored matter fields is affected by the modified gluon propagator in the RGZ environment? These questions are far from having satisfactory answers at the moment and constitute an important topic to be scrutinized in the study of the non-perturbative behavior of correlation functions of colored matter.

In this work, we investigated the impact of the elimination of infinitesimal Gribov copies to the scalar-field propagator. The analysis was carried out by comparing the minimal and non-minimal schemes. As a first step towards the comprehension of the impact of RGZ-propagators in matter loops, we computed the scalar-field propagator at one-loop order in the minimal scheme. In order to compare our findings with the available lattice data for the scalar two-point function, the following strategy was adopted: The mass parameters that are present in the RGZ gluon propagator were obtained from a fitting with lattice simulations for the gluon propagator in pure YM theories. Then, with the parameters fixed, we could predict the scalar-field propagator for different scalar masses. This result was compared with quenched lattice simulations for the scalar field propagator \cite{Maas:2018sqz}. Our findings give a qualitative indication that there is no necessity to introduce a non-minimal coupling between scalars and gauge fields in order to capture the correct infrared behavior of the scalar-field propagator. Those findings seem to be encouraging for further investigations of scalar-gluon vertices in the minimal scheme which, fortunately, also have lattice data available~\cite{Maas:2019tnm}. However, let us stress that investigating further the non-minimal scheme is certainly an important task in order to elucidate better the nature of the coupling between matter and gauge fields when Gribov copies are eliminated. In particular, the dynamical evaluation of the mass parameters present in the non-minimal scheme would provide a self-consistent scenario with no extra fitting parameter. Such issues will be investigated elsewhere.

In order to mitigate the use of the values of the mass parameters of the RGZ propagator obtained from a fitting with different lattice data, we performed a parametric analysis of the impact of variations of those parameters to the scalar propagator together with the gluon propagator. It turns out that the necessary adjustments on the mass parameters to match the lattice data for the scalar propagator severely impact the behavior of the gluon two-point function. Therefore if one aims at quantitative precision in the comparison of the scalar propagator in the deep infrared then further improvements are necessary. The present analysis still faces several limitations that deserve further investigations. From the point of view of the self-consistency of the model, the mass parameters that were obtained through a fitting with the lattice data could be computed from first principles through the solution of their corresponding gap equations. Moreover, in the presence of scalars, one should keep track of their impact to the gluonic dynamics which we just neglected in the present work. Therefore, to some extent, our strategy to fix the mass parameters of the RGZ propagator is very much like a quenched approach. Next to that, we completely neglected the running of those parameters encoded in the running of the coupling $g$, which is crucial for the appropriate match with the deep ultraviolet where standard perturbative YM (with matter) results should be reproduced. Moreover, such running must be introduced in a compatible fashion with our framework, i.e., not displaying Landau poles in the infrared.

We did not find evidence for positivity violation in the scalar sector. As emphasized in the main text, we just investigated sufficient conditions for positivity violation and therefore we cannot draw any strong conclusions on this matter. Yet the lack of evidence for positivity violation was verified in the minimal and non-minimal schemes. This issue certainly deserves further explorations. Moreover, the results here presented suggest that the introduction of a horizon-like function for the matter fields is not necessary in order to reproduce the lattice data qualitatively. Besides avoiding the introduction of extra structures without a clean geometrical motivation as the standard Horizon function, this makes the RGZ-matter model much simpler for computational purposes. However, although we provided a tentative qualitative explanation for justifying such a statement also for quarks, the present work focused on scalars. It is therefore pressing to investigate the more realistic RGZ-QCD model. This will be reported in a future work.

As perspectives for the RGZ-Scalar system, the computation of one-loop vertices is an important step towards the understanding of the self-consistency of the model as well as the evaluation of the propagator by accounting higher loops as done, e.g., in \cite{Pelaez:2017bhh}.

\section*{Acknowledgments}

The authors are grateful to Axel Maas for sharing the lattice data and discussions. ADP is thankful to Reinhard Alkofer for discussions and the University of Graz for hospitality. 
GPB is supported by research grant (29405) from VILLUM fonden.
ADP acknowledges CNPq under the grant PQ-2 (309781/2019-1), FAPERJ under the “Jovem Cientista do Nosso Estado” program (E26/202.800/2019), and NWO under the VENI Grant (VI.Veni.192.109) for financial support. 
PDF acknowledges FAPERJ for the financial support. 

\appendix

\section{Conventions and Feynman Rules}
In this appendix, we present a list of conventions used throughout this paper as well as the Feynman rules of the RGZ-scalar theory in the minimal scheme.

\subsection{Conventions}\label{Sub:AP.Conv}

The generating functional of Euclidean correlation functions $\EuScript{Z}[J]$ is defined as
\begin{equation}
\EuScript{Z}[J] = \int [\EuScript{D}\phi]\,{\rm e}^{-S [\phi] + \int_{x^d} J\phi}\,,
\label{Eq:conv.1}
\end{equation}
with $\phi$ representing a generic collection of fields and $J$ a generic set of external sources coupled to $\phi$. The functional $S[\phi]$ stems for the classical action of the underlying theory. The generating functional of connected correlation functions is defined as
\begin{equation}
\EuScript{W}[J] = {\rm ln}~\EuScript{Z} [J]\,.
\label{Eq:conv.2}
\end{equation}
The classical field or expectation value of $\phi$ at non-vanishing source $\varphi \equiv \langle \phi \rangle_J $ is defined as
\begin{equation}
\varphi (x) = \frac{\delta \EuScript{W}[J]}{\delta J(x)}\,.
\label{Eq:conv.3}
\end{equation}
The generating functional of one-particle irreducible (1PI) correlation functions $\Gamma [\varphi] $ is defined through the Legendre transform,
\begin{equation}
\Gamma [\varphi] = -\EuScript{W}[J_\varphi ] + \int_{x^d} J_{\varphi}\, \varphi (x)\,,
\label{Eq:conv.4}
\end{equation}
with $J_\varphi$ representing the source $J$ written as a functional of the classical field $\varphi$. It follows from \eqref{Eq:conv.4} that
\begin{equation}
\frac{\delta \Gamma [\varphi]}{\delta \varphi (x)} = J_\varphi (x)\,,
\label{Eq:conv.5}
\end{equation}
and
\begin{equation}
\frac{\delta^2 \Gamma [\varphi]}{\delta \varphi (x)\delta \varphi (y)} = \frac{\delta J_\varphi (x)}{\delta \varphi (y)}\,.
\label{Eq:conv.6}
\end{equation}
This leads to
\begin{equation}
\int_{z^d} \frac{\delta^2 \Gamma [\varphi]}{\delta \varphi (x)\delta \varphi (z)}\frac{\delta^2 \EuScript{W} [J]}{\delta J(z) \delta J(y)} = \delta (x-z)\,.
\label{Eq:conv.7}
\end{equation}
Finally, we use the following convention for the Fourier transform for all fields,
\begin{equation}
	\Phi (x) = 
	\int \frac{\mathrm{d}^d p}{(2\pi)^d}\,\mathrm{e}^{-i x\cdot p}\,\tilde{\Phi}(p) \label{ap1} \,.
\end{equation}

\subsection{Feynman Rules}

Tree-level $n$-point vertices $S^{(n)}_{\phi_{i_1} \cdots \phi_{i_n}}$ are defined as 
\begin{equation}
	\begin{aligned}
		&\frac{\delta^n S_\textmd{RGZ}}{\delta \phi_{i_1}(p_1) \cdots \delta \phi_{i_n}(p_n)} \bigg|_{\phi=0} \\
		&\qquad\quad= (2\pi)^d \delta(p_1 + \cdots + p_n) \, \big[ S_{\phi_1 \cdots \phi_n}^{(n)} \big]_{i_1, \cdots , i_n} \,,
	\end{aligned}
\end{equation}
with $i_1, \cdots, i_n$ denoting both Lorentz and SU($N$) indices. 
For the diagrams computed in this paper (c.f.~Fig. \ref{fig:diagrams}), we have used the following vertices
\begin{align}
	[S_{A \phi\phi }]^{abc}_\mu = i g \,f^{abc}\,(p_2 - p_3)_\mu \,,
\end{align}
\begin{align}
	[S_{AA\phi\phi}]^{abcd}_{\mu\nu}= 
	g^2 \delta_{\mu \nu} \left(f^{ace} f^{bde} +  f^{ade}f^{bce} \right) \,,
\end{align}
\begin{align}
	[S_{\phi\phi\phi\phi}]^{abcd} = \frac{\lambda}{3} \left(\delta^{ab} \delta^{cd} + \delta^{ac} \delta^{bd} + \delta^{ad} \delta^{bc}\right) \,.
\end{align}
As for the tree-level scalar and gluon propagators, the expressions are,
\begin{equation}
\langle \phi^a (p) \phi^b (-p)\rangle_0 = \frac{\delta^{ab}}{p^2+m^2_\phi}\,,
\label{Ap:SFPMS.1}
\end{equation}
and
\begin{align}
	\langle A_\mu^a(p) A_\nu^b(-p) \rangle_0 = \delta^{a b} \left(\delta_{\mu \nu} - \frac{p_\mu p_\nu}{p^2}\right) \mathcal{D}_A(p)\,,
	\label{Ap:GP.1}
\end{align}
with the form factor $\mathcal{D}_A(p)$ being
\begin{equation}
	\mathcal{D}_A(p) = \frac{p^2 + M^2}{\left(p^2 + M^2\right) \left(p^2 + m^2\right) + 2\gamma^4 g^2 N}\,.
	\label{Ap:GP.2}
\end{equation}

\bibliography{refs}

\end{document}